\documentclass[10pt,final,doublecolumn]{IEEEtran}

\usepackage{amsmath}
\usepackage{amsthm}
\usepackage{latexsym}
\usepackage{graphicx}
\usepackage{bbding}
\usepackage{indentfirst}
\usepackage{algorithm,algorithmic}
\usepackage{setspace}
\usepackage{float}
\usepackage{epstopdf}
\usepackage{amssymb}
\usepackage{amsfonts}
\usepackage{enumerate}
\usepackage{multicol}
\usepackage{color} 
\usepackage{bm}
\usepackage{amssymb}
\usepackage{stfloats}
\usepackage{epstopdf}
\usepackage{threeparttable}
\usepackage{epstopdf}
\usepackage{threeparttable}
\usepackage{subfigure}
\usepackage{makecell}
\usepackage{cite}
\usepackage{verbatim}
\usepackage{tabularx, booktabs}
\usepackage{array}
\usepackage{enumitem}

\newtheorem{thm}{Theorem}%[section]

\IEEEoverridecommandlockouts
\allowdisplaybreaks[4]
\begin{document}
	\title{Throughput Maximization for Multiuser Communications with Flexible-Sector 6DMA}
	\author{Xiaoming Shi, Yunli Li, \IEEEmembership{Member,~IEEE}, Xiaodan Shao,~\IEEEmembership{Member,~IEEE}, Jie Xu,~\IEEEmembership{Fellow,~IEEE}, and {Rui Zhang},~\IEEEmembership{Fellow,~IEEE}
	\vspace{-1pt}

	\thanks{X. Shi and Y. Li are with School of Science and Engineering, The Chinese University of Hong Kong, Shenzhen, Guangdong 518172, China (e-mail: xiaomingshi@link.cuhk.edu.cn, yunlili@cuhk.edu.cn).}
	\thanks{X. Shao is with Department of Electrical and Computer Engineering, University of Waterloo, Waterloo, ON N2L 3G1, Canada (e-mail: x6shao@uwaterloo.ca).}
	\thanks{J. Xu is with  School of Science and Engineering, the Shenzhen Future Network of Intelligence Institute, and the Guangdong Provincial Key Laboratory of Future Networks of Intelligence, The Chinese University of Hong Kong, Shenzhen, Guangdong, 518172, China (e-mail: xujie@cuhk.edu.cn).}
	\thanks{R. Zhang is with the Department of Electrical and Computer Engineering, National University of Singapore, Singapore 117583 (e-mail: elezhang@nus.edu.sg). }
	}
	
	\maketitle
	\maketitle
	\thispagestyle{empty}

\begin{abstract}
	This paper presents a cost-effective and easily-deployable flexible-sector six-dimensional movable antenna (6DMA) architecture for future wireless communication networks, which enables flexible antenna configurations to match users' spatial distribution for capacity enhancement.
	Different from conventional sectorized base station (BS) with fixed-position antennas (FPAs), the flexible-sector 6DMA-enabled BS employs multiple directional sector antenna arrays that can flexibly move along a common circular track. 
	By properly moving antennas across sectors and rotating all sector antenna arrays synchronously, the flexible-sector BS can adjust the coverage regions of all sectors with flexible antenna allocations over them. 
	In particular, we consider the multiuser downlink communication employing the orthogonal multiple access (OMA) to serve users in each sector. 
	Under this setup, we jointly optimize the sector rotation and the antenna allocation at the flexible-sector BS to maximize the average common throughput achievable for  all users based on their spatial distribution.
	We solve this non-convex optimization problem by deriving closed-form solutions and thereby analyze the effect of users' spatial distribution on the achievable common throughput. 
	It is shown that equal user distribution  over sectors is optimal for maximizing the common throughput.
	Motivated by this result, we further develop a low-complexity suboptimal solution for  the sector rotation that minimizes the variance of user numbers across sectors. 
	Finally, we provide simulation results to verify our analytical results and validate the performance of our proposed solutions. 
	It is demonstrated that the flexible-sector BS significantly improves the network throughput as compared to other benchmark schemes. 
\end{abstract}


\section{Introduction}
	Multiple-input multiple-output (MIMO) technology has played a pivotal role for wireless communication networks, in which base stations (BSs) and user terminals (UTs) are deployed with multiple antennas to enhance the transmission rate and reliability of wireless systems.
	This is achieved by transmitting and receiving multiple data streams simultaneously over the same time-frequency resource, thereby exploiting spatial multiplexing and beamforming gains provided by multiple antennas at the transceivers \cite{1203154,1266912}.
	To further capitalize on these benefits and achieve even higher spatial resolution and more degrees of freedom (DoFs), wireless networks tend to deploy ever-larger antenna arrays at the BSs, leading to the evolution of multiuser MIMO into massive MIMO, and even extremely large-scale MIMO (XL-MIMO) \cite{6736761,9184098,10379539}.
	However, the deployment of large-scale antenna arrays inevitably leads to increasing hardware cost and power consumption.
	To address these challenges, lens-antenna MIMO, holographic MIMO, as well as intelligent reflecting surfaces (IRSs) have been proposed to leverage low-cost materials to achieve high performance gains of MIMO systems, while reducing their implementation costs \cite{7416205,9136592,9326394}.
	Nevertheless, most existing MIMO technologies are based on  conventional fixed-position antennas (FPAs), whose locations are predetermined and remain unchanged once deployed.
	This inherent limitation restricts the ability of MIMO systems to fully exploit the dynamic spatial variations of wireless channels at the transmitter/receiver side.
	
	To overcome the fundamental limitations of FPA systems, movable antenna (MA) \cite{10286328,10318061,11082461} and fluid antenna system (FAS) \cite{9650760,9264694} have emerged as promising solutions.
	By allowing antennas to move freely within a confined region at the BS or UT, MAs/FAS can place antennas to the positions with more favorable channel conditions, thus improving the communication performance.
	Extensive researches have studied MAs/FAS synergizing with other wireless technologies, such as unmanned aerial vehicle (UAV) communications \cite{10654366}, IRS \cite{10539238,10430366}, and integrated sensing and communications (ISAC) \cite{10696953,10839251,11033708}, among others.
	Despite these advancements, existing works on MAs and FAS have mainly focused on antennas' position adjustment while maintaining fixed orientation, which limits their ability to fully exploit the spatial DoFs in a given three-dimensional (3D) space. 
	Moreover, these systems are mostly designed for mitigating small-scale channel fading based on instantaneous channel state information (CSI).
	As a result, their achievable  performance is fundamentally constrained by the quality and timeliness of channel estimation, as well as practical hardware limitations such as the speed of antenna movement.
	
	To address these challenges and achieve full flexibility in antenna deployment at the BS, six-dimensional movable antenna (6DMA) has been proposed as an effective solution for further improving wireless network capacity \cite{10945745,10883029}.
	At 6DMA-enabled BSs, multiple antenna arrays are connected to a central processing unit (CPU) and can each be independently adjusted in both 3D position and 3D rotation (orientation) \cite{10752873,10848372}.
	The available movement space for each 6DMA array depends on the site size and hardware implementation, within which it can move and rotate in 3D space flexibly  using extendable and rotatable arms  \cite{10752873,10848372}.
	To achieve low implementation complexity and reduce the practical hardware change of existing BSs from conventional FPA arrays to fully adjustable 6DMA arrays, antenna elements or arrays can also move over the surface of a sphere \cite{hua205}, or along a circular track \cite{ming,ming2} via motor-driven mechanisms, thus simultaneously adjusting their rotations (orientations) and positions efficiently.

	While these studies demonstrate the effectiveness of 6DMA-enabled BS designs for capacity enhancement, they cannot be applied to existing BSs without changing their antenna architecture (i.e., sector antenna arrays) and working principle (i.e., sectorized cell coverage). 
	To address this practical implementation difficulty while still achieving flexible antenna movement and allocation over sectors of existing BSs, this paper presents an easily implementable and deployable 6DMA-enabled BS architecture,  termed the flexible-sector BS, where  multiple sector antenna arrays are mounted along a common circular track.
	As shown in Fig. \ref{draft_fig1}, by rotating all sector antenna arrays in a synchronized manner via rotating the circular track as well as moving antenna elements between the antenna arrays of adjacent sectors along the circular track, the flexible-sector BS can dynamically adjust the orthogonal coverage regions of all sectors with flexible antenna allocations over them, so as to effectively match the users' spatial distribution and thereby improve their achievable rates. 
	Mathematically, the flexible-sector BS architecture can be considered as a special case of the general 6DMA model. 
	However, this practical design preserves the essential spatial flexibility needed to reconfigure sector coverage and antenna allocation, while substantially reducing the 6DMA  implementation complexity.
	Moreover, its simplified structure leads to more tractable  system  and channel models as compared to the general 6DMA, thereby facilitating efficient performance analysis and design optimization. 
	The main contributions of this paper are summarized as follows:
	\begin{itemize}
		\item First, we present the flexible-sector BS architecture with multiple directional sector antenna arrays, which supports flexible antenna allocation and synchronized sector rotation. 
		Based on this architecture, we characterize the channel model in terms of the  sector rotation and antenna allocation, and derive the transmit power required at the BS to achieve a given transmission rate for each user by considering the orthogonal multiple access (OMA).
		Moreover, we extend the conventional homogeneous Poisson point process (HPPP) to model the spatially non-uniform user distribution in the angular domain.
		Based on the knowledge of user distribution, we further derive a closed-form expression for the total average transmit power required for the BS to achieve the same transmission rate for all users over all sectors (termed as common throughput). Different from the approach of ergodic rate approximation using  Monte Carlo simulations in previous 6DMA studies (e.g., \cite{10752873,10848372}), the analytical results derived in this paper greatly facilitate the performance evaluation and design optimization for the flexible-sector BS system.
		\item Second, we formulate an optimization problem to maximize the common throughput achievable for all users by jointly designing the sector rotation and the antenna allocation across sectors, subject to the discrete rotation constraint and the total budgets of antennas and transmit power at the BS.
		To solve this non-convex optimization problem, we adopt a bisection search to iteratively determine the maximum achievable common throughput.
		In each iteration, we perform a one-dimensional search over discrete sector rotation values and optimize the antenna allocation with the fixed sector rotation by using the Lagrange multiplier method.
		\item Third, to investigate the effect of users' spatial distribution on the system throughput and draw key insights into the optimal design of sector rotation (which essentially changes the user distributions over sectors), we consider an ideal case where the users can be arbitrarily allocated over sectors provided that their total number equals a given value. 
		Then, we optimize the user allocations over sectors to maximize the common throughput either jointly with antenna allocation or under fixed antenna allocation.
		It is shown that under joint optimization, the common throughput is maximized when the number of users is equal over all sectors, whereas under fixed antenna allocation, the optimal user distribution exhibits a water-filling structure.
		Based on these results, we derive a theoretical upper bound for the common throughput and propose an alternative low-complexity solution for the sector rotation to maximize the common throughput.
		\item Finally, we provide simulation results to validate the performance of the flexible-sector BS system and  optimization algorithms.
		Compared to the conventional fixed-sector BS, the proposed scheme significantly improves the achievable system throughput by effectively adjusting user distribution across sectors via sector rotation as well as flexibly allocating antennas over sectors via antenna movement.
	\end{itemize}
	
	The rest of this paper is organized as follows.
	Section \ref{II} introduces the flexible-sector BS architecture and the system model.
	Section \ref{III} formulates the common throughput maximization problem.
	Section \ref{IV} presents an efficient algorithm to solve the formulated problem.
	Section \ref{V} investigates  the effect of user distribution across sectors on the maximum common throughput and presents a low-complexity algorithm for solving  the considered problem.
	Section \ref{VI} provides simulation results for performance evaluation and comparison.
	Finally, Section \ref{VII} concludes this paper.

	\textit{Notations}: $a$, $\bf a$, $\bf A$, and $\mathcal{A}$ denote a scalar, a vector, a matrix, and a set, respectively.
	$(\cdot)^T$ denotes the transpose.
	$\mathbb{E}[\cdot]$ denotes the expected value of random variable.
	$[{\bf a}]_n$ denotes the $n$-th entry of vector $\bf a$.
	$[{\bf A}]_{i,j}$ denotes the entry in row $i$ and column $j$ of matrix $\bf A$.
	${\rm Var}(\bf a)$ denotes the variance of the entries in vector $\bf a$.
	$|\mathcal{A}|$ represents the cardinality of set $\mathcal{A}$.
	${\bf 0}_t $ denotes the $t \times 1$ all-zero vector.
	$\mathcal{A}\cap\mathcal{B}$ represents the intersection set of $\mathcal{A}$ and $\mathcal{B}$.
	$\mathbb{R}$, $\mathbb{R}_+$, and $\mathbb{R}_{++}$ denote the sets of real numbers, non-negative real numbers, and positive real numbers, respectively.
	$\mathbb{Z}$, $\mathbb{Z}_+$, and $\mathbb{Z}_{++}$ denote the sets of integers, non-negative integers, and positive integers, respectively.
	For any set $\mathcal{X}$, $\mathcal{X}^{M\times N}$ denotes the set of all $M \times N$ matrices whose entries belong to $\mathcal{X}$ (column vectors when $N=1$).
	$a\mod b$ denotes the modulo operation.
	$\lfloor \cdot \rfloor$ denotes the floor operator.
	$\max \left\{\cdot\right\}$ and $\min \left\{\cdot\right\}$ denote the selection of the maximum and minimum values, respectively.
	
		%${\mathbb{S}}^{M\times N}$ denotes the set of all $M\times N$ matrices (or $M$-dimensional column vectors for $N=1$) with entries in set ${\mathbb{S}}$.
	\vspace{-2 pt}
	\section{System Model}\label{II}

	\subsection{Flexible-Sector BS Architecture}
	
	\begin{figure}[t]
		\centering
		\includegraphics[width=3.5in]{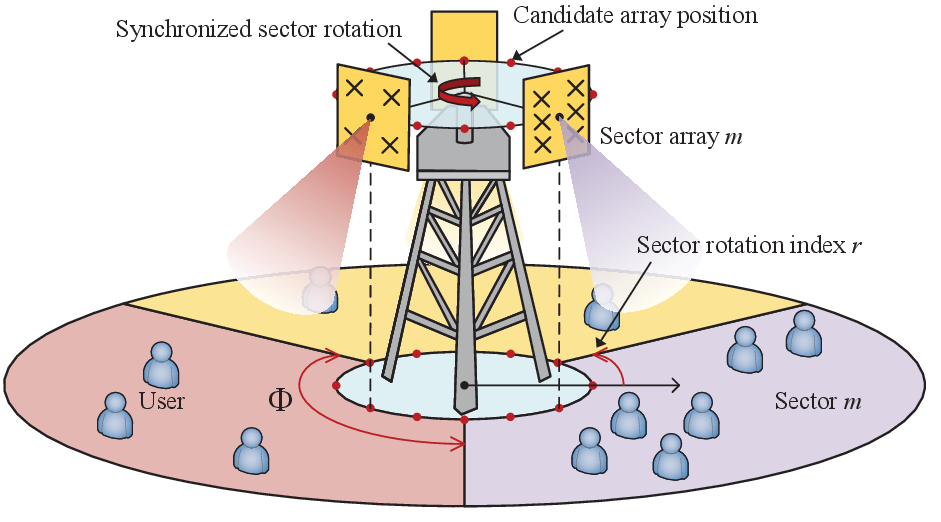}
		\caption{Illustration of the flexible-sector BS.}
		\label{draft_fig1}
	\end{figure}
	
	We consider a downlink communication system where one BS serves multiple single-antenna users distributed in its coverage area.
	As shown in Fig. \ref{draft_fig1}, the BS is equipped with $M\in\mathbb{Z}_{++}$ sector antenna arrays which are equally placed along a circular track parallel to the ground.
	Each sector antenna array is responsible for covering one sector of the BS's service area,  denoted by the set $\mathcal{M}=\{1,2,\cdots,M\}$.
	In addition, each sector has an equal azimuth coverage angle, denoted as $\Phi = \frac{2\pi}{M}$.
	The azimuth angular center of each sector is aligned with the center of its corresponding antenna array. 
	Moreover, each sector antenna array consists of $N_m$ directional antennas, $m\in\mathcal{M}$, denoted by a collective vector ${\bf n}=[N_1,N_2,\cdots,N_M]^T$, with $\sum_{m=1}^{M} N_m = N$, where $N$ denotes the total number of antennas at the BS and $N \geq M$. 
	We assume $N_m\in\mathbb{Z}_{++},\forall m\in\mathcal{M},$ to guarantee that each sector has at least one antenna.
	Different from the conventional BS with a fixed and equal number of antennas across all sectors, we consider that the antennas at the BS can be flexibly allocated over different sectors.
	Specifically, the antennas can move freely along the circular track from one sector to another such that the number of antennas in each sector antenna array can be flexibly adjusted.

\begin{figure}[t]
		\centering
		\includegraphics[width=2.7in]{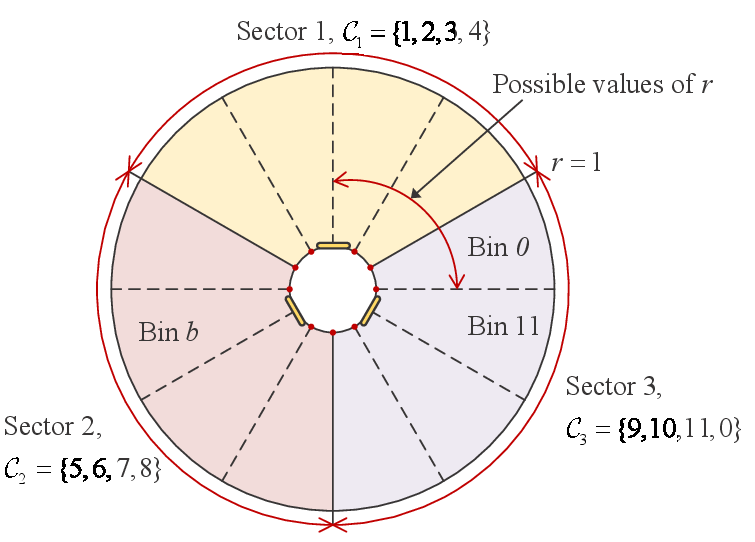}\vspace{-0 pt}
		\caption{Illustration of the sector division when $M=3$, $L=4$, $B=12$, and $r=1$.}
		\label{draft_fig2}
	\end{figure}

	To further enhance flexibility, all sector antenna arrays can rotate horizontally in a synchronized manner while maintaining a fixed relative angular difference of $\Phi$ between any two adjacent sector arrays, thus leading to different and yet orthogonal coverage regions of all sectors in the azimuth angle via rotation.
	We set the circular track's center as the reference position of the BS, and define it as the pole of a polar coordinate system.
	For ease of practical implementation, we consider that the candidate positions of all sector antenna arrays are constrained to $B=LM$ discrete ones with  $L\in\mathbb{Z}_{++}$, which are equally spaced along the circular track with a rotation step $\delta=\frac{2\pi}{B}$.
	Correspondingly, the coverage area of the BS can be equally partitioned into $B$ orthogonal bins in terms of the azimuth angle, denoted by the set $\mathcal{B}=\{0,1,\cdots,B-1\}$, each with an angular interval $\left[{b\delta, (b+1)\delta}\right)$, $b\in\mathcal{B}$ (see Fig. \ref{draft_fig2}).
	Due to the discrete sector rotation, each sector can always cover $L$ adjacent bins.
	We define the sector rotation as the index of the bin where the first sector edge is encountered in the counter-clockwise direction from the polar axis, denoted by the integer $r$, with $0\leq r \leq L-1$.\footnote{The bins covered by different sectors undergo a cyclic shift when $r$ exceeds $L-1$.}
	The set of bins covered by sector $m$ is thus given by
	\begin{equation}\label{C_m}
		\begin{aligned}
		\mathcal{C}_m({r})=&\{b\in\mathcal{B} \mid b = (\omega_m(r)+l)\!\mod B,\\
		&\quad\quad\quad\quad l=0,1,\cdots,L-1\},\quad m\in\mathcal{M},
		\end{aligned}
	\end{equation}
	where $\omega_m({r}) = r+(m-1)L$ denotes the starting bin index of sector $m$.
		
	\vspace{-2 pt}
	\subsection{Channel Model}
	\vspace{-2 pt}
	We denote the set of users in the BS's coverage area as $\mathcal{K} = \cup_{m=1}^M\mathcal{K}_m(r)$, where $\mathcal{K}_m(r)$ denotes the set of users located in sector $m$, depending on the sector rotation $r$.
	For analytical tractability, we assume that each sector has at least one user regardless of the value of $r$.
	To avoid interference across different sectors, the antennas in each sector are assumed to adopt a directional radiation pattern (in the case of $M > 1$).
	Specifically, we consider a normalized and truncated radiation pattern, for which the total radiation power over the whole space is equal to one \cite{balanis2016antenna}.
	Given that the azimuth beamwidth of each antenna is set to $\Phi$, the effective gain of each antenna at sector $m\in\mathcal{M}$ to user $k$ is given by
	\begin{equation}\label{G}\vspace{-2 pt}
		G_k(r,m) = \left\{ {\begin{array}{*{20}{l}}
				{ \dfrac{2\pi}{\Phi} ,}&{\text{if } k\in\mathcal{K}_m(r),} \\ 
				0 ,&{\text{otherwise,}} 
		\end{array}} \right.
		\quad k\in\mathcal{K}.
	\end{equation}
	
	We assume fading channels between the BS and users, each constituting the distance-dependent path gain and an additional random term accounting for small-scale fading.
	Specifically, given user $k$ at a horizontal distance $d_k\in[D_{\rm min},D_{\rm max}]$ from the reference position of the BS, the path gain for user $k$ is given by 
	\begin{equation}\label{h_k}\vspace{-2 pt}
		h_k = {\beta_0}{\left(d_k^2+H^2\right)^{-\frac{\alpha_0}{2}}},\quad k\in\mathcal{K},
	\end{equation}
	where $D_{\rm min}$ and $D_{\rm max}$ represent the minimum and maximum coverage radii of the BS, respectively, $\alpha_0$ denotes the path loss exponent, $H$ denotes the height of the reference position of the BS, and $\beta_0 = (\frac{4\pi f_c}{c})^{-2}$ denotes the channel power gain at a reference distance of $1$ meter ($\rm m$) from the BS with $f_c$ denoting the carrier frequency and $c$ denoting the speed of light.
	Based on \eqref{G} and \eqref{h_k} as well as the fact that $\Phi = \frac{2\pi}{M}$, the channel power gain for user $k\in\mathcal{K}_m(r)$ is modeled as
	\begin{equation}\label{ChannelPowerGain}\vspace{-2 pt}
			g_k(r,N_m) =   N_m M h_k \varsigma_k,\quad k\in\mathcal{K}_m(r),m\in\mathcal{M},
	\end{equation}
	where $N_m$  accounts for the array gain (by assuming maximal ratio transmission for the antennas in each sector), $M$  accounts for the directional antenna gain in each sector, and $\varsigma_k \sim {\text{Exp}}(1)$ denotes an independent and identically distributed (i.i.d.) exponential random variable with unit mean accounting for the small-scale channel fading (assumed to be Rayleigh fading).
	Note that in \eqref{ChannelPowerGain}, the sector rotation $r$ determines the mapping of users to their respective covering sectors and thus the array gain $N_m$ for each user.
	Accordingly, $g_k$ is a function of $r$, even though $r$ does not explicitly appear in its expression.
	
	\subsection{Achievable Rate}
	We consider the OMA for the downlink data transmissions from the BS to different users.
	Specifically,  all sectors can reuse the total available time and frequency resources as there is no inter-sector interference due to the use of directional antennas at all sectors.
	Within each sector, users are served in orthogonal resource blocks (RBs) for avoiding the intra-sector interference.
	Assume that the BS allocates transmit power $p_k$ to communicate with user $k\in\mathcal{K}_m(r)$ in sector $m\in\mathcal{M}$.
	Then, the achievable rate of user $k$ in bits per second per Hz (bps/Hz) is expressed as 
	\begin{equation}\label{R_k}\vspace{-2 pt}
		\begin{aligned}
		R_k(r,N_m) &= \frac{1}{K_m(r)}\log_2\left(1+\frac{p_k g_k(r,N_m)K_m(r)}{\sigma^2 }\right)\\
		& = \frac{1}{K_m(r)}\log_2\left(1+\frac{p_k N_m M h_k \varsigma_k {K_m(r)}}{\sigma^2}\right),\\
		&\quad\quad\quad\quad\quad\quad\quad\quad\  k\in\mathcal{K}_m(r),m\in\mathcal{M},
		\end{aligned}
	\end{equation}
	where %$\eta_0 = \frac{\pi\beta_0}{\Phi\sigma^2}$, 
	$K_m(r) = \left| {\mathcal{K}_m(r)} \right|$ denotes the number of users in sector $m$, and  the receiver noise is assumed to be additive white Gaussian with average power $\sigma^2$.
	
	Denote the common (minimum) throughput achievable for all users in all sectors as $\bar{R}$ in bps/Hz.
	Due to the small-scale channel fading, an outage event occurs when the BS-user channel  cannot support the desired common throughput $\bar{R}$.
	According to \eqref{R_k} and the cumulative distribution function (CDF) of exponential random variable $\varsigma_k$, the outage probability for user $k$ is given by
	\begin{equation}
	\begin{aligned}
	{\text{P}_{{\rm out},k}}(r,N_m) & = \mathbb{P}\left\{R_k(r,N_m)<\bar{R}\right\}\\
	& = \mathbb{P}\left\{ \varsigma_k < \frac{\sigma^2\left(2^{{K_m(r)}\bar{R}}-1\right)}{p_k N_m M h_k {K_m(r)}}  \right\}\\
	& = 1-{\exp}\left(-\frac{\sigma^2\left(2^{{K_m(r)}\bar{R}}-1\right)}{p_k N_m M h_k {K_m(r)}} \right),\\
	&\quad\quad\quad\quad\quad\quad\  k\in\mathcal{K}_m(r),m\in\mathcal{M}.
	\end{aligned}
	\end{equation}
	Denote the allowed maximum outage probability as ${\bar{\text{P}}}_{\rm out}$ for all BS-user links.
	Then, the transmit power required to achieve the common throughput $\bar{R}$, while satisfying the outage probability requirement ${\bar{\text{P}}}_{\rm out}$ for user $k$, is derived from the outage probability constraint ${\text{P}_{{\rm out},k}}(r,N_m) \leq {\bar{\text{P}}}_{\rm out}$, given by  
	\begin{equation}\label{p_k}
		\begin{aligned}
			%p_k(r,N_m,\bar{R}) &= \frac{2^{{K_m(r)}\bar{R}}-1}{\eta_0 N_m {K_m(r)}} \left(d^2+H^2\right)^{\frac{\alpha_0}{2}},\\
			p_k(r,N_m,\bar{R}) &\geq \frac{\sigma^2}{\tau  N_m M h_k {K_m(r)}} \left(2^{{K_m(r)}\bar{R}}-1\right),\\
			&\quad\quad\quad\quad\quad\quad\quad\quad k\in\mathcal{K}_m(r),m\in\mathcal{M},
		\end{aligned}
	\end{equation}
	where $\tau = -\ln(1-{\bar{\text{P}}}_{\rm out})$.
	\subsection{User Distribution}\label{II-D}
	To model the users' spatial distribution, we assume that the distribution of users in each bin $b\in\mathcal{B}$ follows an independent HPPP $\mathcal{P}_b$ with density $\rho_b > 0$ (users/m$^2$). 
	Specifically, users in each bin are independent and uniformly distributed with the number of users in each bin following a Poisson distribution, denoted as \cite{8368129}
	\begin{equation}\label{K_b}
	{\hat K}_b \sim {\text{Poisson}} \left({\hat\lambda}_b\right),\quad b\in\mathcal{B},
	\end{equation} 
	where ${\hat\lambda}_b = \frac{\pi\left(D_{\max}^2-D_{\min}^2\right)}{B}\rho_b$ denotes the average number of users in bin $b$.
	This model assumes local homogeneity within each bin with a small angular span $\delta$.
	In addition, $\rho_b$ can be different across bins to capture the non-uniform users' spatial distribution in practice.
	Due to the independence of Poisson random variables $\hat{K}_b$, $b\in\mathcal{B}$, the number of users in sector $m\in\mathcal{M}$,  denoted as $K_m(r) = \sum_{b\in\mathcal{C}_m(r)}{\hat{K}_b}$, also follows a Poisson distribution, i.e., $K_m(r) \sim {\text{Poisson}} \left(\lambda_m(r)\right)$, where $\lambda_m(r)$ denotes the average number of users in sector $m$, given by \cite{Haenggibook}
	\begin{equation}\label{barK_m}
		\lambda_m(r)  = \sum_{b\in\mathcal{C}_m(r)} {\hat\lambda_b},\quad m\in\mathcal{M}.
	\end{equation}

	\section{Problem Formulation}\label{III}
	Based on the users' spatial distribution and the transmit power required to achieve the common throughput $\bar{R}$ given by \eqref{p_k}, the minimum total average transmit power required for sector $m$ is given by  
	\begin{equation}\label{barp_m}
		\bar{p}_m(r,N_m,\bar{R}) = \mathbb{E}_{\mathcal{P}_m}\left[\sum_{k\in\mathcal{K}_{m}(r)}p_k(r,N_m,\bar{R})\right],~m\in\mathcal{M},
	\end{equation}
	where the expectation is with respect to (w.r.t.) the users' spatial distribution $\mathcal{P}_m=\{\mathcal{P}_b\}_{b\in\mathcal{C}_m(r)}$, i.e., their random horizontal distances from the BS, and the random number of users in sector $m$, $K_m(r)$.
	Define $\bar{\gamma} (\bar{R})= 2^{\bar R}-1$ as the signal-to-noise ratio (SNR) required to achieve $\bar{R}$.
	The minimum total average transmit power required for sector $m\in\mathcal{M}$ in \eqref{barp_m} is expressed in a closed-form expression, shown in the following theorem.
	\begin{thm}\label{thm1}
		The minimum total average transmit power required for sector $m$ is  
		\begin{equation}\label{p_m}
				\bar{p}_m(r,N_m,\bar{R}) = \frac{1}{\gamma_m(N_m)}e^{\bar{\gamma}(\bar{R})\lambda_m(r)},\quad m\in\mathcal{M} ,
		\end{equation}
		where $\gamma_m(N_m)$ denotes the average SNR in sector $m$, given by
		\begin{equation}\label{gamma_m}
			\gamma_m(N_m) = \frac{\tau M N_m}{\nu_0\sigma^2},
		\end{equation}
		with
		\begin{equation}
			\nu_0 = \frac{\left(D_{\max}^2+H^2\right)^{\frac{\alpha_0}{2}+1}-\left(D_{\min}^2+H^2\right)^{\frac{\alpha_0}{2}+1}}{\beta_0\left(D_{\max}^2-D_{\min}^2\right)\left(\frac{\alpha_0}{2}+1\right)}.
		\end{equation}
	\begin{IEEEproof}
			See Appendix A.
	\end{IEEEproof}
	\end{thm}
	From Theorem \ref{thm1}, it follows that the required average transmit power per sector $\bar{p}_m,m\in\mathcal{M},$ is inversely proportional to the number of antennas $N_m$ in sector $m$, but increases exponentially with the average number of users $\lambda_m(r)$ in it. 
	Also note that $\lambda_m(r)$ is related to the sector rotation $r$.
	Thus, it is revealed that $\bar{p}_m$'s can be varied across sectors by adjusting $N_m$'s and $r$. 
%	This indicates that the unbalanced load among sectors will lead to even bigger unbalance in power consumption, which ultimately increase the total transmit power required at the BS. 
	Let $P_{\max}$ denote the total average transmit power at the BS.
	We thus have the following power constraint:
	\begin{equation}
		\sum_{m=1}^{M}\bar{p}_m(r,N_m,\bar{R}) \leq P_{\max}.
	\end{equation}

	In this paper, we aim to maximize the common throughput $\bar{R}$ by jointly optimizing the sector rotation $r$ and the antenna allocation $\bf n$ across sectors.
	Accordingly, the optimization problem is formulated as 
	\begin{align}
		\underset{r\in\mathbb{Z}_{+},\mathbf{n} \in \mathbb{Z}_{++}^{M \times 1},\bar{R}\in\mathbb{R}_+}{\max} \quad & \bar{R} \label{P1}\\
		\text{s.t.}\quad  &   0\leq r \leq L-1, \tag{\ref{P1}{a}} \label{P1con1} \\
		& \sum_{m=1}^{M} N_m \leq N, \tag{\ref{P1}{b}} \label{P1con2} \\
		%&  N_m\in\mathbb{Z}_{+},\quad \forall m\in\mathcal{M} \tag{\ref{P1}{c}}  \label{P1con3} \\
		%& \bar{p}_m(r,N_m,\bar{R})\geq 0,\quad \forall m\in\mathcal{M}, \tag{\ref{P1}{c}} \label{P1con3} \\
		& \sum_{m=1}^{M} \bar{p}_m(r,N_m,\bar{R})\leq P_{\max}, \tag{\ref{P1}{c}} \label{P1con4}
	\end{align}
	where constraints \eqref{P1con2} and \eqref{P1con4} are due to the total budgets of the antennas and the transmit power at the BS, respectively.
	We denote the optimal solution to problem \eqref{P1} as $\left(r^*,{\bf n}^*\right)$ and the corresponding maximum common throughput as $\bar{R}^*$.
	Problem \eqref{P1} is difficult to solve as it is a mixed-integer non-convex  optimization problem. 
	This is because variables $r$ and $\bf n$ are discrete, and constraint \eqref{P1con4} is non-convex  due to the fact that $\lambda_m(r)$ in \eqref{p_m} is not a convex function w.r.t. $r$.
	 %coupling between $\lambda_m(r)$ and $r$, which introduces non-convexity in 
	Nevertheless, we solve problem \eqref{P1} efficiently in the next section.

	\section{Proposed Algorithm for Solving Problem \eqref{P1}}\label{IV}
	%Solving problem \eqref{P1} is non-trivial due to the discrete variables $r$ and $\bf n$, and the non-convex constraint \eqref{P1con4}.
	%By exploiting its special structure, problem \eqref{P1} can be optimally solved as follows.
	Problem \eqref{P1} can be equivalently solved by considering a series of sub-problems given below, each for a given $\bar{R}$.
	Furthermore, $\bar{R}$ can be updated via the bisection search method until its convergence is attained.
	Specifically, to check if a given $\bar{R}$ is achievable, we solve the following sub-problem that minimizes the total average transmit power at the BS subject to constraints \eqref{P1con1} and \eqref{P1con2}, i.e.,
	\begin{align}
		\underset{r\in\mathbb{Z}_{+},\mathbf{n} \in  {\mathbb{Z}_{++}^{M \times 1}} }{\min} \quad & \sum_{m=1}^{M} \bar{p}_m(r,N_m,\bar{R}) \label{P2}\\
		\text{s.t.}\quad  &  \eqref{P1con1},~\eqref{P1con2}.\nonumber
	\end{align}
	If the optimal value of problem \eqref{P2} is no larger than $P_{\max}$, then constraint \eqref{P1con4} is satisfied, and the optimal solution to problem \eqref{P2} together with its corresponding $\bar{R}$ becomes a feasible solution to problem \eqref{P1}.
	On the other hand, if the optimal value of problem \eqref{P2} is larger than $P_{\max}$, then the corresponding $\bar{R}$ is not achievable.
	Accordingly, bisection search can be applied to find the maximum common throughput $\bar{R}^*$ for problem \eqref{P1} iteratively. 
	Thus, we focus on solving problem \eqref{P2} in the following for a given $\bar{R}$, or equivalently with given  $\bar{\gamma}=2^{\bar R}-1$.
	
	Note that problem \eqref{P2} is an integer programming problem and can be optimally solved by using an exhaustive search method (ESM) over all possible combinations of $r$ and $N_m,m\in\mathcal{M},$ within their respective discrete sets.
	The computational complexity of the ESM is of order $\mathcal{O}\left( L C_{N-1}^{M-1} \right)$, where $C_{N-1}^{M-1}$ denotes the binomial coefficient, representing the number of possible antenna allocation combinations among $M$ sectors with $N$ antennas. 
	However, this method can be computationally prohibitive as either $N$ or $M$ increases, and moreover, it cannot reveal any analytical insight into the structure of the optimal solution.
	Thus, in this paper, we solve problem \eqref{P2} by adopting a two-step approximation method.
	Specifically, we first fix $r$ and relax the integer constraints on $N_m,m\in\mathcal{M},$ into their continuous counterparts to derive the optimal antenna allocation ${\bf n}^*$ with given $r$ and $\bar{R}$.
	Then,  we perform a one-dimensional search over all discrete values of  $r$  to obtain its solution for problem \eqref{P2}. 
	This is feasible since the candidate set for sector rotation $r$ is finite and discrete.
	
	\subsection{Antenna Allocation Optimization}\label{IV.A}
	Given the fixed sector rotation $r$ and the target common throughput $\bar R$, problem \eqref{P2} reduces to optimizing the antenna allocation $\bf n$.
	To address the combinatorial complexity introduced by the integer constraints $N_m \in\mathbb{Z}_{++},\ m\in\mathcal{M}$, we relax each $N_m$ to a non-negative real number, and denote the relaxed antenna allocation as ${\hat{\bf n}} = [\hat{N}_1,\hat{N}_2,\cdots,\hat{N}_M]^T\in{\mathbb{R}_+^{M\times 1}}$.
	The resulting problem for optimizing ${\hat{\bf n}}$ is formulated as
	\begin{align}
		\underset{{\hat{\bf n}}\in{\mathbb{R}_+^{M\times 1}}}{\min} \quad & \sum_{m=1}^{M} \frac{\zeta_m}{\hat{N}_m} \label{P3}\\
		\text{s.t.}\quad  &  \sum_{m=1}^{M} \hat{N}_m \leq N , \tag{\ref{P3}{a}} \label{P3con1}%\\
	%	& \hat{N}_m\geq 1,\quad \forall m \in\mathcal{M}, \tag{\ref{P3}{b}} \label{P3con2}
	\end{align}
	where $\zeta_m = \frac{\nu_0\sigma^2}{\tau M}e^{\bar{\gamma}\lambda_m(r)}$ is fixed and independent of ${\hat{\bf n}}$.
	It can be verified that problem \eqref{P3} is a convex optimization problem because its objective is convex w.r.t. ${\hat{N}_m},\forall m\in\mathcal{M}$, and the inequality constraint \eqref{P3con1} is linear.
	To solve problem \eqref{P3}, we employ the Lagrange multiplier method.
	%Note that we temporarily ignore the inequality constraints \eqref{P3con2}, which can simplify the mathematical formulation of the Lagrange multiplier method and enables the derivation of closed-form solutions for the relaxed problem. 
	%And constraints \eqref{P3con2} can be satisfied later through a rounding and redistribution process.
	The Lagrangian for the relaxed problem \eqref{P3} is given by 
	\begin{equation}
		\mathcal{L}\left({\hat{\bf n}},\mu\right) = \sum_{m=1}^{M} \frac{\zeta_m}{\hat{N}_m}+\mu\left(\sum_{m=1}^{M} \hat{N}_m - N \right),
	\end{equation}
	where $\mu \geq 0$ is the Lagrange multiplier corresponding to the inequality constraint \eqref{P3con1}.
	By taking the first derivative of ${\mathcal{L}\left({\hat{\bf n}},\mu\right)}$  w.r.t. $\hat{N}_m$ and setting it equal to zero, we have
	\begin{equation}\label{partialN_m}
		\begin{aligned}
		\frac{{\partial \mathcal{L}\left({\hat{\bf n}},\mu\right)}}{{\partial {\hat{N}_m}}} = - \frac{\zeta_m}{\hat{N}_m^2} + \mu = 0\quad\Rightarrow\quad & \hat{N}_m = \sqrt{\frac{\zeta_m}{\mu}},\quad m\in\mathcal{M}.
		\end{aligned}
	\end{equation}
	By substituting \eqref{partialN_m} into constraint \eqref{P3con1}, we have
	\begin{equation}
		\sqrt{\mu} = \frac{1}{N}\sum_{m=1}^{M}\sqrt{\zeta_m}.
	\end{equation}
	Thus, the optimal continuous antenna allocation solution to problem \eqref{P3} with given $r$ and $\bar R$ is given by 
	\begin{equation}
	\begin{aligned}\label{hatNm_opt}
		\hat{N}_m^* &= \frac{N\sqrt{\zeta_m}}{\sum_{m'=1}^{M}\sqrt{\zeta_{m'}}}\\
		& = \frac{N{ e^{\frac{\bar{\gamma}}{2}\lambda_m(r)}}}{\sum_{m'=1}^{M}{e^{\frac{\bar{\gamma}}{2}\lambda_{m'}(r)}}},\quad m\in\mathcal{M}.
	\end{aligned}
	\end{equation}
	It follows from \eqref{hatNm_opt} that the optimal number of antennas allocated to sector $m$, $\hat{N}_m^*$, is proportional to ${e^{\frac{\bar{\gamma}}{2}\lambda_m(r)}}$. 
	Thus, the sector with a larger average number of users $\lambda_m$ should be allocated with more antennas for enhancing the array gain and thereby reducing its average transmit power, so as to minimize the total transmit power at the BS. 
	%In addition, the square root dependence implies diminishing returns in allocating additional antennas to a sector. 
	%Specially, for sector $m$ with heavy user load $\bar{K}_{m}\bar{R}\gg 1$, the numerator term can be approximated as $\sqrt{2^{\bar{K}_{m}\bar{R}}-1}\approx 2^{\bar{K}_{m}\bar{R}/2}$.
	%The exponent $\bar{K}_{m}\bar{R}$ is halved, suppressing the exponential resource dominance.
	%	This prevents over-allocation to heavily loaded sectors at the expense of others, achieving the fairness.
	If all sectors have identical average number of users, i.e., $\lambda_m = \lambda_{m'}, \forall m,m'\in\mathcal{M},m\neq m'$, the optimal antenna allocation in \eqref{hatNm_opt} reduces to the equal antenna allocation over sectors, i.e.,  $\hat{N}_m^* = \frac{N}{M},\forall m\in\mathcal{M}$.
	
	To obtain integer solutions for antenna allocation $N_m \in\mathbb{Z}_{++},m\in\mathcal{M}$, for problem \eqref{P2}, we modify the optimal  continuous solution derived in \eqref{hatNm_opt} as follows.
	%This approach can obtain a suboptimal but feasible integer antenna allocation.
	First, we enforce the constraints $N_m\geq 1,m\in\mathcal{M},$ by pre-allocating one antenna for each sector.
	Then, we allocate the remaining $N-M$ antennas over all sectors according to \eqref{hatNm_opt}.
	Thus, given fixed $r$ and $\bar R$, the integer antenna allocation solution to problem \eqref{P2} is given by
	\begin{equation}\label{integer_ant_allo}
		{N}_m^* = 1+\left\lfloor  \frac{(N-M){e^{\frac{\bar{\gamma}}{2}\lambda_m(r)}}}{\sum_{m'=1}^M{e^{\frac{\bar{\gamma}}{2}\lambda_{m'}(r)}}}  \right\rfloor ,\quad m\in\mathcal{M}.
	\end{equation}
	Note that the above solution is always feasible to problem \eqref{P2}, but in general only a suboptimal solution  given finite values of $N$ and $M$.
	Nevertheless, for practical setups with large values of $N$ with $N\gg M$, it can be verified by simulation that the solution in \eqref{integer_ant_allo} is very close to the optimal solution of problem \eqref{P2} via the ESM (see Section \ref{VI}).

	\subsection{Sector Rotation Optimization}
	%The continuous antenna allocation given in \eqref{hatNm_opt} provides a close approximation of the actual integer antenna allocation while maintaining proportionality to user load in each sector.
	Based on the integer antenna allocation solution in \eqref{integer_ant_allo}, we can obtain the total average transmit power at the BS by substituting ${N}_m^*,m\in\mathcal{M},$ into the objective of problem \eqref{P2}, yielding 
	\begin{equation}\label{P_tot}
		P_{\rm tot}(r) = \frac{\nu_0\sigma^2}{\tau M}\sum_{m=1}^M \frac{e^{\bar{\gamma}\lambda_m(r)}}{{N}_m^*(r)}.
	\end{equation}
	Thus, the optimization problem w.r.t. sector rotation $r$ is formulated as
	\begin{align}
		\underset{r\in\mathbb{Z}_{+}}{\min} \quad & P_{\rm tot}(r) \label{P4}\\
		\text{s.t.}\quad  &  \eqref{P1con1}.\nonumber
	\end{align}
	%It can be observed from problem \eqref{P4} that the expected number of users in each sector $\lambda_{m},m\in\mathcal{M},$ as given in \eqref{barK_m}, changes discontinuously with sector rotation $r$ due to the coverage mapping $\mathcal{C}_m({r})$ in \eqref{C_m}.
	%In addition, the floor operation in $N_m^*$ exhibits step-wise discontinuities.
	%Thus, problem \eqref{P4} is highly non-convex and combinatorial over the integer-valued $r$.
	Since the feasible set of $r$ given in \eqref{P1con1} consists of a finite number of discrete values, problem \eqref{P4} can be simply solved by a one-dimensional search over all possible values of $r$. 
	
	Let $\bar{R}_{\rm up}$ and $\bar{R}_{\rm low}$ denote the upper and lower bounds of the common throughput $\bar{R}$, respectively, and $\epsilon$ denote a small positive constant that controls the accuracy of the bisection search method for updating $\bar{R}$.
	The overall algorithm for solving problem \eqref{P1} is summarized in Algorithm 1.
	The computational complexity of this algorithm can be shown to be of order $\mathcal{O}(LM\log_2\frac{\bar{R}_{\rm up} - \bar{R}_{\rm low}}{\epsilon})$, where the logarithmic term accounts for the number of bisection iterations required to achieve  accuracy $\epsilon$.
	Note that this complexity is much lower than that of the ESM for solving problem \eqref{P1}.  
	
	\begin{algorithm}[!t]
		\renewcommand{\thealgorithm}{1:}
		\caption{The overall algorithm for solving problem \eqref{P1}}\label{alg:alg1}
		\begin{algorithmic}[1]
			\STATE \textbf{Input:} $N$, $M$, $L$, $P_{\max}$, ${\bar{\text{P}}}_{\rm out}$, $\sigma^2$, $D_{\max}$, $D_{\min}$, $\alpha_0$, $f_c$, $\bar{R}_{\rm up}$, $\bar{R}_{\rm low}$, $\{\hat{\lambda}_b\}_{b\in\mathcal{B}}$, $\epsilon$.
			\STATE \textbf{Output:} $r^*$, ${\bf n}^*$, $\bar{R}^*$.
			\STATE \textbf{Initialization:} Set $r^* = 0$, ${\bf n}^* = {\bf 0}_M$, and $\bar{R}^* = 0$.
			\STATE {\textbf{while}} $\bar{R}_{\rm up}-\bar{R}_{\rm low}> \epsilon$ {\textbf{do}} 
			\STATE \hspace{0.2cm} Set $\bar{R} = \frac{1}{2}(\bar{R}_{\rm up} + \bar{R}_{\rm low})$.
			\STATE \hspace{0.2cm} Set $P_{\rm tot}^* = +\infty$.
			\STATE \hspace{0.2cm} {\textbf{for}} $r=0:1:L-1$ {\textbf{do}} 
			\STATE \hspace{0.5cm} Initialize ${\bf n}^{(r)} = {\bf 0}_M$.
			\STATE \hspace{0.5cm} {\textbf{for}} $m=1:1:M$ {\textbf{do}}
			\STATE \hspace{0.7cm} Calculate average number of users $\lambda_{m}(r)$ in \eqref{barK_m}.
			\STATE \hspace{0.5cm} {\textbf{end for}}
			\STATE \hspace{0.5cm} {\textbf{for}} $m=1:1:M$ {\textbf{do}}
			\STATE \hspace{0.7cm} Calculate integer antenna allocation $N^*_m(r)$ in \eqref{integer_ant_allo}.
			\STATE \hspace{0.7cm} $[{\bf n}^{(r)}]_m \leftarrow N^*_m(r)$.
			\STATE \hspace{0.5cm} {\textbf{end for}}
			\STATE \hspace{0.5cm} Calculate total average transmit power $P_{\rm tot}(r)$ in \eqref{P_tot}.
			\STATE \hspace{0.5cm} {\textbf{if}} $P_{\rm tot}(r)<P_{\rm tot}^*$ {\textbf{then}}
			\STATE \hspace{0.7cm} $P_{\rm tot}^* \leftarrow P_{\rm tot}(r)$.
			\STATE \hspace{0.7cm} ${\bf n}^* \leftarrow {\bf n}^{(r)}$.
			\STATE \hspace{0.7cm} $r^* \leftarrow r$.
			\STATE \hspace{0.5cm} {\textbf{end if}}
			\STATE \hspace{0.2cm} {\textbf{end for}}
			\STATE \hspace{0.2cm} {\textbf{if}} $P_{\rm tot}^* \leq P_{\max}$ {\textbf{then}}
			\STATE \hspace{0.5cm} $\bar{R}_{\rm low} \leftarrow \bar{R}$.
			\STATE \hspace{0.5cm} $\bar{R}^* \leftarrow \bar{R}$.
			\STATE \hspace{0.2cm} {\textbf{else}} 
			\STATE \hspace{0.5cm} $\bar{R}_{\rm up} \leftarrow \bar{R}$.
			\STATE \hspace{0.2cm} {\textbf{end if}} 
			\STATE {\textbf{end while}}
			\STATE \textbf{return} $r^*$, ${\bf n}^*$, $\bar{R}^*$.
		\end{algorithmic}
		\label{alg1}
	\end{algorithm}
	
	\subsection{Special Cases: Flexible Antenna Allocation or Sector Rotation Only} \label{IV.C}
	In practice, the joint optimization of sector rotation and antenna allocation may not be implementable due to hardware limitations and complexity considerations.
	Therefore, in this subsection, we present algorithms for obtaining the solutions of problem \eqref{P1} for the cases of flexible antenna allocation only and flexible sector rotation only, respectively.  
	
	\subsubsection{Case 1: Antenna Allocation Optimization Only with Fixed Sector Rotation}
	When $r$ is fixed (e.g., $r = 0$), antenna allocation  $\bf n$ can be  optimized for each target common throughput $\bar{R}$ according to \eqref{integer_ant_allo}. 
	In this case, we can implement Algorithm 1 by omitting the search over $r$ (i.e., lines 7–22).
	For each $\bar{R}$, ${\bf n}^*$ is directly computed via \eqref{integer_ant_allo}, and the total average transmit power is evaluated using \eqref{P_tot}.
	Bisection search over $\bar{R}$ is still implemented  until the desired accuracy $\epsilon$ is achieved. 
	The complexity order of this simplified algorithm reduces to $\mathcal{O}(M\log_2\frac{\bar{R}_{\rm up} - \bar{R}_{\rm low}}{\epsilon})$.
	\subsubsection{Case 2: Sector Rotation Optimization Only with Fixed Antenna Allocation}
	When $\bf n$ is fixed (e.g., equal antenna allocation over sectors with  $N_m = \lfloor N/M \rfloor,\forall m\in\mathcal{M}$), we can still implement Algorithm 1 to solve problem \eqref{P1} by replacing the antenna allocation update steps (i.e., lines 12-15) with the  equal antenna allocation given fixed  $N_m$.
	Although the order of the algorithm complexity remains the same as that of Algorithm 1, the actual algorithm running time is significantly reduced since the antenna allocation computation in \eqref{integer_ant_allo} is omitted.
	
	\section{Effects of User Distribution on Maximum Common Throughput}\label{V}
	
	In the previous section, the optimization of sector rotation $r$ helps adjust the user distribution across sectors, i.e., the average number of users $\lambda_{m}$ for sector $m\in\mathcal{M}$, by rotating the coverage regions $\mathcal{C}_m(r)$'s of all sectors.
	However, the effects of changing  the user distribution or $\lambda_{m}$'s across sectors by sector rotation $r$ on the maximum system common throughput remain unexplored.
	To fill this gap and gain insights, we consider in this section an ideal setup, where $\lambda_{m}$'s across sectors can be arbitrarily adjusted subject to the constraint on their given sum (or the total average number of users served by the BS).
	Under this setup, we maximize common throughput $\bar{R}$ by considering two cases when $\lambda_{m}$'s are optimized jointly with antenna allocation $\bf n$ and when antenna allocation $\bf n$ is fixed and can be unequal over sectors, respectively. 
	% We show that in the former case, $\bar{R}$ is maximized when $\lambda_{m}$'s are equal across all sectors and accordingly derive a theoretically upper bound for $\bar{R}$. 
	Inspired by the analytical results, we further propose an alternative solution to problem \eqref{P2}, which has a lower computational complexity than the proposed solution algorithm in Section \ref{IV}. 
	%Then, in the latter case,  we investigate the optimization of user distribution under a fixed antenna allocation over sectors (which can be unequal over sectors) to reveal the corresponding optimal user distribution that maximizes $\bar{R}$ and draw more useful insights.
	
	\subsection{Joint User Distribution and Antenna Allocation Optimization}\label{V-A}
	
	Let $\boldsymbol{\lambda} = [\lambda_1,\lambda_2,\cdots,\lambda_M]^T \in{\mathbb{R}_{++}^{M\times 1}}$ denote the collective vector constituting the average  numbers of users over all sectors. %\footnote{Here, we assume that $\lambda_m$ is independent of $r$, as the user distribution across sectors can be arbitrarily allocated.}
	Similar to problem \eqref{P2}, we aim to minimize the total average transmit power over all sectors to meet a given common throughput $\bar{R}$. 
	However, different from problem \eqref{P2} that jointly optimizes sector rotation $r$ and antenna allocation $\bf n$, in this subsection, we consider a new problem that jointly optimizes $\boldsymbol{\lambda}$ and $\bf n$. 
	% Similar to the procedure in Section \ref{IV}, we use a bisection search method over common throughput $\bar{R}$ and formulate a corresponding power minimization problem for each candidate $\bar{R}$ as formulated in problem \eqref{P2}.
	For ease of analysis, we relax the integer antenna allocation $\bf n$ to its continuous counterpart $\hat{\bf n}$ as in problem \eqref{P3}. 
	As such, this optimization problem  is formulated as
	\begin{align}
		\underset{{\boldsymbol{ \lambda} }\in{\mathbb{R}_{++}^{M\times 1}},{\hat{\bf n}}\in{\mathbb{R}_+^{M\times 1}}}{\min} \quad & \sum_{m=1}^{M} \frac{1}{\gamma_m({\hat N}_m)}e^{\bar{\gamma}\lambda_m} \label{P5}\\
		\text{s.t.}\quad  &  \sum_{m=1}^{M} \lambda_m = \lambda_{\rm sum}, \tag{\ref{P5}{a}} \label{P5con1}\\
		&  \sum_{m=1}^{M} \hat{N}_m \leq N , \tag{\ref{P5}{b}} \label{P5con2}
	\end{align}
	where $\gamma_m({\hat N}_m)$ is given in \eqref{gamma_m}, and $\lambda_{\rm sum} = \sum_{b=0}^{B-1} {\hat\lambda_b}$ denotes the total average number of users in the whole coverage area of the BS.
	
	For any given $\boldsymbol{\lambda}$, problem \eqref{P5} reduces to problem \eqref{P3} w.r.t. $\hat{\bf n}$.
	Thus, the optimal continuous antenna allocation ${\hat N}_m^*$ across all sectors can be expressed as the closed-form solution in \eqref{hatNm_opt}.
	By substituting ${\hat N}_m^*$ into the objective function of problem \eqref{P5}, the total average transmit power with optimal continuous antenna allocation is given by
	\begin{equation}\label{hatP_tot}
		{\hat P}_{\rm tot}({\boldsymbol{ \lambda} }) =  \frac{\nu_0\sigma^2}{\tau M N}\left(\sum_{m=1}^{M} e^{\frac{\bar{\gamma}}{2}\lambda_m} \right)^2.
	\end{equation}
	Thus, problem \eqref{P5} is simplified to minimizing the summation of the exponential terms in \eqref{hatP_tot}, i.e., 
	\begin{align}
		\underset{{\boldsymbol{ \lambda} }\in{\mathbb{R}_{++}^{M\times 1}}}{\min} \quad & \sum_{m=1}^{M} e^{\frac{\bar{\gamma}}{2}\lambda_m} \label{P6}\\
		\text{s.t.}\quad  &  \eqref{P5con1}. \nonumber
	\end{align}
	The objective function of problem \eqref{P6} is the summation of $M$ convex exponential functions of identical forms. 
	Therefore, it is easy to show by Jensen's inequality that the optimal solution to problem \eqref{P6} is given by  
	%a convex problem as the objective function is convex due to the summation of exponential terms, and the feasible region is a convex set defined by the linear equality constraint \eqref{P5con1} and inequality constraints $\lambda_{m}> 0,\forall m \in\mathcal{M}$.
%	It can be efficiently solved to obtain the global optimum by using the Lagrange multiplier method.
%	The Lagrangian with multiplier $\hat{\mu} \geq 0$ corresponding to the equality constraint \eqref{P5con1} is given by
%	\begin{equation}
%		{\hat{\mathcal{L}}}({\boldsymbol{ \lambda}},\hat{\mu}) = \sum_{m=1}^{M} e^{\frac{\bar{\gamma}}{2}\lambda_m}   - \hat{\mu}\left( \sum_{m=1}^{M} \lambda_m -  \lambda_{\rm sum} \right).
%	\end{equation}
%	The derivative of the Lagrangian w.r.t. $\lambda_{m}$ is given by
%	\begin{equation}\label{partialmu1}
%		\frac{{\partial {\hat{\mathcal{L}}}({\boldsymbol{ \lambda}},\hat{\mu})}}{\partial \lambda_m} = {\frac{{\bar \gamma}}{2}}e^{\frac{\bar{\gamma}}{2}\lambda_m}-\hat{\mu}, \quad m\in\mathcal{M}.
%	\end{equation}
%	By setting \eqref{partialmu1} equal to zero, we have 
%	\begin{equation}\label{lambda_m1}
%		\lambda_m = \frac{2}{\bar{\gamma}}\ln \frac{2\hat{\mu}}{\bar{\gamma}}.
%	\end{equation}
%	By substituting \eqref{lambda_m1} into constraint \eqref{P5con1}, the optimal user distribution is given by
	\begin{equation}\label{optlambda1}
		\lambda_m^* = \frac{\lambda_{\rm sum}}{M},\quad \forall m\in\mathcal{M}.
	\end{equation}
	From \eqref{optlambda1}, it follows that the optimal user distribution across sectors for the total average power minimization should lead to equal average number of users in all sectors. 
	Accordingly, the corresponding optimal continuous antenna allocation over sectors is also equal, i.e., 
	\begin{equation}\label{optn1} 
		\hat{N}_m^* = \frac{N}{M}, \quad m \in \mathcal{M}. 
	\end{equation}

%	This is because the exponential power scaling $e^{\frac{\bar{\gamma}}{2}\lambda_{m}}$ fundamentally dominates the system performance.
%	While the optimal antenna allocation compensates for the imbalance of user distribution, it cannot mitigate the exponentially increasing power consumption when any sector is overloaded.
%	The sector rotation provides a spatial degree of freedom (DoF) that dynamically realigns the coverage region $\mathcal{C}_m(r)$ for each sector, thereby changing the real-world user's spatial distribution for each sector toward the equalized distribution.
	
	Based on \eqref{hatP_tot} and \eqref{optlambda1}, we obtain a lower bound for the total average transmit power at the BS required to achieve  a given $\bar{R}$ as
	\begin{equation}\label{powerlowerbound}
		P_{\rm tot}^{\rm L} = \frac{ M\nu_0\sigma^2}{\tau N}e^{\frac{\bar{\gamma}\lambda_{\rm sum}}{M}}.
	\end{equation}
	According to \eqref{powerlowerbound} and the total power constraint \eqref{P1con4}, an upper bound for the maximum achievable common throughput $\bar{R}$ is obtained when $P_{\rm tot}^{\rm L}$ is equal to $P_{\max}$, i.e., 
	\begin{equation}\label{Rupperbound}
		\bar{R}^{\rm U} = \log_2\left( 1 + \frac{M}{\lambda_{\rm sum}}\ln\frac{P_{\max} N \tau}{M\nu_0\sigma^2}\right).
	\end{equation}
	Note that under the assumptions of continuous antenna allocation and arbitrarily adjustable user distribution, the optimal value of problem \eqref{P1} can achieve the upper bound $\bar{R}^{\rm U}$ in \eqref{Rupperbound}.
	However, in practice, since adjusting the sector rotation generally cannot equalize the average numbers of users $\lambda_m$'s across sectors, the optimal value of problem \eqref{P1} may not reach this upper bound.
	
%	Note that $\bar{R}^{\rm U}$ given in \eqref{Rupperbound} is also an upper bound for the optimal value of problem \eqref{P1} under the assumptions of continuous antenna allocation and arbitrarily adjustable user distribution.
%	Since by adjusting sector rotation $r$, the resulting average numbers of users $\lambda_m$'s across sectors may not be equal for practical user distribution, the optimal value of problem \eqref{P1} may not achieve that given in \eqref{Rupperbound}.    
	
	\subsection{User Distribution Optimization with Fixed Antenna Allocation}
	%With the fixed antenna allocation, the common throughput maximization reduces to optimizing the user distribution, i.e., $\boldsymbol{\lambda}$.
	In the previous subsection, we consider the joint optimization of user distribution $\boldsymbol{\lambda}$ and antenna allocation $\bf n$. 
	In contrast, in this subsection, we consider the optimization of $\boldsymbol{\lambda}$ only by assuming that $\bf n$ is given and fixed. 
	This scenario is relevant in practical systems where the antenna array is predetermined due to hardware limitations or sector planning constraints.
	For given $\bar{R}$ and $\bf n$, the total average transmit power minimization problem \eqref{P5} is simplified to 
	\begin{align}
		\underset{{\boldsymbol{ \lambda} }\in{\mathbb{R}_{++}^{M\times 1}}}{\min} \quad & \sum_{m=1}^{M} \frac{1}{\gamma_m}e^{{\bar \gamma}\lambda_m} \label{P7}\\
		\text{s.t.}\quad  &  \eqref{P5con1},\nonumber
	\end{align}
	where $\gamma_m$ is a constant since $N_m$ is fixed in each sector.
	Problem \eqref{P7} is a convex optimization problem since its objective is the summation of $M$  convex functions, and the feasible region is a convex set defined by the linear equality constraint \eqref{P5con1} and inequality constraints $\lambda_{m}> 0,\forall m \in\mathcal{M}$.
	By using the Lagrange multiplier method, we show that the optimal user distribution across sectors to minimize the total average transmit power should correspond to a water-filling solution based on the average number of users allocated in different sectors.
	Specifically, by letting  $\tilde{\mu} \geq 0$ denote the Lagrange multiplier corresponding to the equality constraint \eqref{P5con1}, the Lagrangian for problem \eqref{P7} is given by	
	\begin{equation}
		{\tilde{\mathcal{L}}}({\boldsymbol{ \lambda}},\tilde{\mu}) = \sum_{m=1}^{M} \frac{1}{\gamma_m}e^{{\bar \gamma}\lambda_m}  - \tilde{\mu}\left( \sum_{m=1}^{M} \lambda_m -  \lambda_{\rm sum} \right).
	\end{equation}
	The first derivative of the Lagrangian w.r.t. $\lambda_{m}$ is given by
	\begin{equation}\label{partialmu}
		\frac{{\partial {\tilde{\mathcal{L}}}({\boldsymbol{ \lambda}},\tilde{\mu})}}{\partial \lambda_m} = \frac{{\bar \gamma}}{\gamma_m}e^{{\bar \gamma}\lambda_m}-\tilde{\mu}, \quad m\in\mathcal{M}.
	\end{equation}
	By setting \eqref{partialmu} equal to zero, the optimal user allocation is given by
	\begin{equation}\label{optlambda}
		\lambda_m^* =  \left\{ {\begin{array}{*{20}{l}}
				{\frac{1}{\bar{\gamma}}\left(\ln\frac{\tilde{\mu}}{\bar{\gamma}} - \ln \frac{1}{\gamma_m}\right),  }&{\text{if } \gamma_m \geq \frac{\bar{\gamma}}{\tilde{\mu}},} \\ 
				{0,}&{\text{otherwise,}} 
		\end{array}} \right.  \quad m\in\mathcal{M},
	\end{equation}
	where $\tilde{\mu}$ should be chosen such that the constraint for total average number of users is met with equality, i.e.,
	\begin{equation}
		\sum_{m=1}^{M} \max\left\{\ln\frac{\tilde{\mu}}{\bar{\gamma}} - \ln \frac{1}{\gamma_m},0\right\} = \bar{\gamma}\lambda_{\rm sum}.
	\end{equation}
	From \eqref{optlambda}, the optimal user distribution is to take advantage of sectors with larger average SNR $\gamma_m$ (due to larger array gain $N_m$ with more antennas) by assigning more users to them.
%	For the sector with small $\gamma_m$, less users are allocated to this sector.
%	If $\gamma_m$ falls below the cutoff value $\frac{\bar{\gamma}}{\tilde{\mu}}$, this sector is not used.
	For the special case when the antennas are equally allocated to $M$ sectors, i.e., $\gamma_m(N_m) = \frac{\tau N}{\nu_0\sigma^2},\forall m \in\mathcal{M}$, \eqref{optlambda} reduces to an equal user allocation across sectors, i.e.,  $\lambda_m^* = \frac{\lambda_{\rm sum}}{M}$, which is the same as that in \eqref{optlambda1}, as expected.
	
	\subsection{Low-Complexity Algorithm for Solving Problem \eqref{P2} Based on User Number Variance Minimization }\label{V-C}
%	The above theoretical analysis demonstrates that, under optimal joint user distribution and antenna allocations, equalizing $\lambda_{m},m\in\mathcal{M},$ across sectors maximizes the common throughput.
	The optimality of equal user distribution discussed in Section~\ref{V-A} implies that a lower variance in average numbers of users $\lambda_{m}$'s among sectors leads to better performance in terms of the common throughput.
	Motivated by this, we propose an alternative sector rotation design that aims to minimize the variance of $\lambda_{m}$'s across sectors.
	Specifically, instead of calculating the optimal antenna allocation and the resulting total average transmit power in Algorithm 1 for every candidate sector rotation $r$, we calculate the entry variance among $\lambda_{m}$'s across sectors in vector $\boldsymbol{\lambda}(r) = [\lambda_1(r), \ldots, \lambda_M(r)]^T$ for each $r$, defined as $\operatorname{Var}(\boldsymbol{\lambda}(r))$, and select the optimal sector rotation $r^*$ achieving the minimum variance, i.e., 
	\begin{equation}
		r^* = \arg\min_{0\leq r \leq L-1}~\operatorname{Var}(\boldsymbol{\lambda}(r)).
	\end{equation}
	After the optimal $r^*$ is determined, the corresponding optimal antenna allocation $\bf n^*$ is calculated only once for the given  $r^*$ using  \eqref{integer_ant_allo}.
	The above solution to problem \eqref{P2} is summarized in Algorithm 2.
	Similarly, by replacing lines 7–22 for solving problem \eqref{P2} in Algorithm~1 with Algorithm 2, a lower-complexity alternative algorithm for solving problem \eqref{P1} is also obtained.  

	\begin{algorithm}[!t]
		\renewcommand{\thealgorithm}{2:}
		\caption{The low-complexity algorithm for solving problem \eqref{P2}}\label{alg:alg2}
		\begin{algorithmic}[1]
			\STATE \textbf{Input:} $N$, $M$, $L$, $\{\hat{\lambda}_b\}_{b\in\mathcal{B}}$, $\bar{R}$.
			\STATE \textbf{Output:} $r^*$, ${\bf n}^*$, $P_{\rm tot}^*$.
			\STATE \textbf{Initialization:} Set ${\bf n}^* = {\bf 0}_M$ and $v^{(r)} =  0,\forall r$.
			\STATE {\textbf{for}} $r=0:1:L-1$ {\textbf{do}} 
			\STATE \hspace{0.3cm} Initialize ${\boldsymbol{ \lambda} }^{(r)} = {\bf 0}_M$.
			\STATE \hspace{0.3cm} {\textbf{for}} $m=1:1:M$ {\textbf{do}}
			\STATE \hspace{0.7cm} Calculate average number of users $\lambda_{m}(r)$ in \eqref{barK_m}.
			\STATE \hspace{0.7cm} $[{\boldsymbol{ \lambda} }^{(r)}]_m \leftarrow \lambda_{m}(r)$.
			\STATE \hspace{0.3cm} {\textbf{end for}}
			\STATE \hspace{0.3cm} $v^{(r)} \leftarrow  \operatorname{Var}(\boldsymbol{\lambda}^{(r)})$.
			\STATE {\textbf{end for}}
			\STATE $r^* = \arg\min_{0\leq r \leq L-1}~v^{(r)} $.
			\STATE  {\textbf{for}} $m=1:1:M$ {\textbf{do}}
			\STATE \hspace{0.3cm} Calculate integer antenna allocation $N^*_m(r^*)$ in \eqref{integer_ant_allo}.
			\STATE \hspace{0.3cm} $[{\bf n}^*]_m \leftarrow N^*_m(r^*)$.
			\STATE  {\textbf{end for}}
			\STATE  Calculate total average transmit power $P_{\rm tot}(r^*)$ in \eqref{P_tot}.
			\STATE  $P_{\rm tot}^* \leftarrow P_{\rm tot}(r^*)$.
			\STATE \textbf{return} $r^*$, ${\bf n}^*$, $P_{\rm tot}^*$.
		\end{algorithmic}
		\label{alg2}
	\end{algorithm}
	
	 The computational complexity of Algorithm 2 is of order $\mathcal{O}(LM)$, which is the same as that for solving problem \eqref{P2} in Algorithm 1.
	 Despite having the same complexity order, the actual computation time of Algorithm 2 with finite $L$ and $M$ is significantly reduced, since the optimal antenna allocation and resulting total average transmit power are computed only once.
%	 Furthermore, the variance computation across different rotations can be parallelized, enhancing efficiency. 
%	 The proposed low-complexity algorithm achieves comparable performance with much lower implementation cost, faster execution, and better scalability for large-scale or real-time systems.
	
	\section{Simulation Results}\label{VI}
	In this section, we provide simulation results to validate our analysis and evaluate the performance of flexible-sector BS-enabled communication systems.
	\subsection{Simulation Setup and Benchmark Schemes}
	In the simulation, the maximum and minimum coverage radii of the BS are set as $D_{\max}=100~{\rm m}$ and $D_{\min}=20~{\rm m}$, respectively, with the height of BS's reference position given by $H=20~{\rm m}$.
	The coverage area is partitioned into $M=3$ sectors, each further divided into $L=10$ angular bins, resulting in $B=30$  orthogonal bins in total.
	The BS is equipped with a total of $N=300$ antennas, and is subject to a maximum transmit power constraint of $P_{\max} = 40~{\rm dBm}$.
	The path loss exponent is set as $\alpha_0 = 2.5$.
	The system operates at a carrier frequency of $f_c = 24~{\rm GHz}$.
	The noise power is set as $\sigma^2 = -114~{\rm dBm}$.
	A maximum outage probability of ${\bar{\text{P}}}_{\rm out}=0.01$ is required for all served users. 
	For the proposed algorithm, the initial interval for bisection search over $\bar{R}$ is set as $\bar{R}_{\rm up}=\bar{R}^{\rm U}$ given in \eqref{Rupperbound} and $\bar{R}_{\rm low}=0$, and the search accuracy is specified by $\epsilon = 10^{-4}$.

	For the non-uniform user distribution, the average number of users ${\hat{\lambda}}_b$ in each bin $b\in\mathcal{B}$ is configured  to capture the realistic distributions of both background and hotspot users.
	Specifically, ${\hat\lambda_b}$ is defined as the sum of the average number of background users (that are assumed to be uniformly distributed in all bins of the cell) in each bin and the average number of users contributed by all hotspot areas in bin $b$, given by
	\begin{equation} 
		{\hat\lambda_b} = {\hat\lambda}_{\mathrm{bg}} + \sum_{i=1}^{N_\mathrm{hot}}{\hat\lambda}_{\mathrm{hot},i} \mathbb{I}(b \in \mathcal{H}_i), \quad b\in\mathcal{B},
	\end{equation}
	where ${\hat\lambda}_{\mathrm{bg}}$ denotes the average number of background users per bin, ${N_\mathrm{hot}}$ denotes the number of hotspot areas, and ${\hat\lambda}_{\mathrm{hot},i}$ denotes the additional average number of users per bin contributed by hotspot area $i\in\{1,\cdots,{N_\mathrm{hot}}\}$.
	The indicator function $\mathbb{I}(b \in \mathcal{H}_i)$ equals $1$ if bin $b$ falls within the coverage of hotspot area $i$, and $0$ otherwise, where $\mathcal{H}_i$ denotes the set of bin indices covered by hotspot area $i$, corresponding to its span and location.
	For hotspot area $i$, its starting bin index and span are denoted as $\omega_{\mathrm{hot},i}\in\mathcal{B}$ and  $L_{\mathrm{hot},i}$, respectively.
	The set of bins covered by hotspot area $i$ is thus defined as
	\begin{equation}
		\begin{aligned}
			\mathcal{H}_i=&\{b\in\mathcal{B} \mid b = (\omega_{\mathrm{hot},i}+l)\!\mod B,\\
			&\quad l=0,1,\cdots,L_{\mathrm{hot},i}-1\},\quad i=1,\cdots,N_\mathrm{hot}.
		\end{aligned}
	\end{equation}
	We set the total average number of users in the whole coverage area of the BS as $\lambda_{\rm sum} = 100$, and consider $N_\mathrm{hot}=3$ hotspot areas with the average number of background users per bin specified as ${\hat\lambda}_{\mathrm{bg}}=1$.
	The additional average numbers of users contributed by the three hotspot areas follow the ratio ${\hat\lambda}_{\mathrm{hot},1}:{\hat\lambda}_{\mathrm{hot},2}:{\hat\lambda}_{\mathrm{hot},3} = 1:2:4$, and their starting bin indices and spans are set as ${\left[\omega_{\mathrm{hot},1},\omega_{\mathrm{hot},2},\omega_{\mathrm{hot},3}\right]}={\left[0,3,15\right]} $ and $[L_{\mathrm{hot},1},L_{\mathrm{hot},2},L_{\mathrm{hot},3}]=[3,4,6]$, respectively. 
%	To capture diverse user distribution scenarios, the starting bin indices of all hotspot areas $\omega_{\mathrm{hot},i},i=1,\cdots,N_\mathrm{hot},$  are independent and uniformly selected from the set of bins $\mathcal{B}$, i.e., $\omega_{\mathrm{hot},i} \sim \mathrm{Unif}\{0,1,\cdots,B-1\}$.
	
	In addition to the proposed joint antenna allocation and sector rotation optimization scheme, three benchmark schemes are defined for performance comparison as follows:
	\begin{itemize}
		\item \textbf{Antenna allocation optimization only with fixed sector rotation:}
		This scheme considers a fixed sector rotation $r=0$, and applies the simplified algorithm described in Special Case 1 of Section \ref{IV.C} to optimize the antenna allocation across sectors only.
	\end{itemize}
	\begin{itemize}
		\item \textbf{Sector rotation optimization only with fixed antenna allocation:} 
		This scheme considers a fixed and equal antenna allocation over sectors, i.e., $N_m = \left\lfloor N/M \right\rfloor$, $\forall m\in\mathcal{M}$, and applies  the simplified algorithm described in Special Case 2 of Section \ref{IV.C} to optimize sector rotation $r$ only.
	\end{itemize}
	\begin{itemize}
		\item \textbf{Fixed antenna allocation and sector rotation:}
		This scheme considers the conventional fixed-sector BS system, where antennas are equally allocated across sectors, i.e., $N_m = \left\lfloor N/M \right\rfloor$, $\forall m\in\mathcal{M}$, and sector rotation is fixed and set as $r = 0$.
	\end{itemize}
	
%	\begin{figure*}[t]	
%		\centering
%		\subfigure[Average transmit power per sector versus expected user number.]{\includegraphics[width=0.45\textwidth]{Verify_theorem1.eps}}
%		\subfigure[Average transmit power per sector versus antenna number per sector.]{\includegraphics[width=0.45\textwidth]{Verify_theorem1_2.eps}}
%		\caption{Verify the accuracy of theorem 1.}
%		\label{result1}
%	\end{figure*}
\subsection{Validation of Analytical Results and Algorithm Effectiveness}
	\begin{figure}[t]
		\centering
		\includegraphics[width=0.45\textwidth]{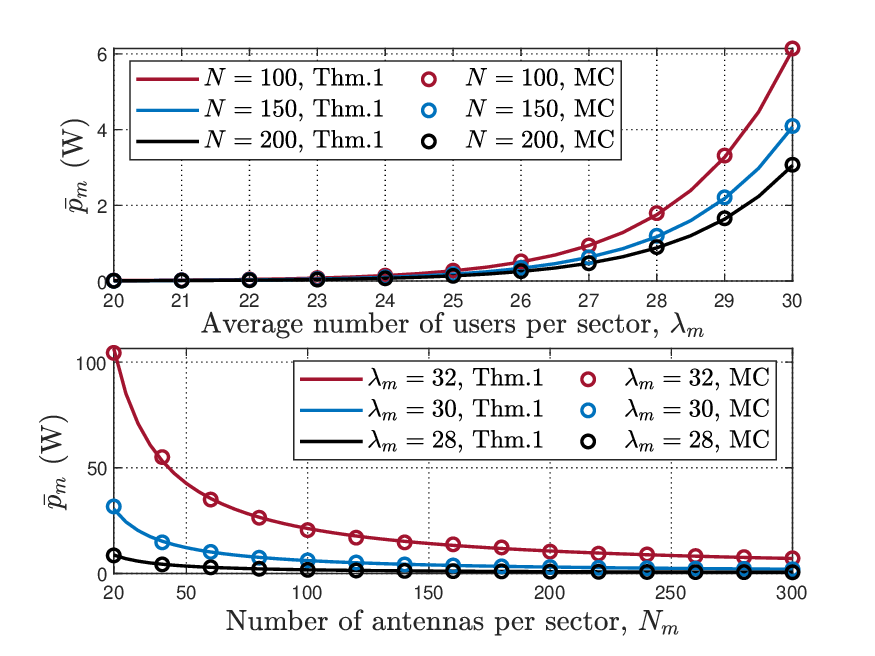}
		\caption{Validation  of Theorem 1 with $\bar{R}=0.7~{\rm bps/Hz}$.}
		\label{result1}
	\end{figure}
	
	Fig. \ref{result1} presents a comparison between the total average transmit power for a representative sector $m$ derived in Theorem \ref{thm1} and that obtained via Monte Carlo (MC) simulations with a uniform user distribution, i.e., $N_{\rm hot} = 0$ and $\lambda_m = \lambda_{m'}, \forall m,m'\in\mathcal{M},m\neq m'$.
	Since all sectors are statistically equivalent under this setting, the transmit power performance for one sector is representative of the entire system.
	The MC simulations are performed as follows.
	For each value of the average number of users in sector $m$, $\lambda_{m}$, we generate $10^7$ independent realizations.
	For each realization, the actual number of users in this sector, $K_m$, is sampled from $K_m \sim {\text{Poisson}} \left(\lambda_m\right)$.
	The user locations are randomly generated within this sector according to the spatial distribution specified in Section \ref{II-D}.
	For each user, the required transmit power to achieve the fixed common throughput $\bar{R}$ is computed according to \eqref{p_k}.
	The total transmit power for this sector is then obtained as the sum of transmit power for all users, and the sample mean over all realizations is computed as the MC result.
	As shown in Fig. \ref{result1}, the analytic curves from Theorem \ref{thm1} perfectly match with the MC simulation results, which verifies the accuracy of the derived expressions.
	In addition, it is observed that $\bar{p}_m$ increases exponentially with $\lambda_{m}$ (in the top sub-figure) as a result of serving a larger number of users simultaneously, and decreases inversely with $N_m$ (in the bottom sub-figure) due to the increasing array gain.
	
	\begin{figure}[t]
		\centering
		\includegraphics[width=0.45\textwidth]{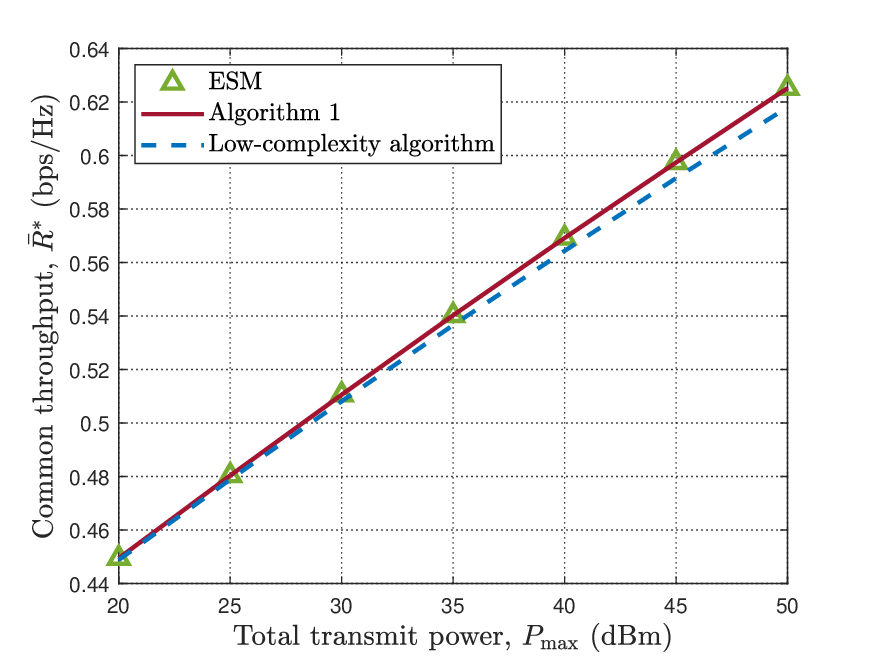}
		\caption{Maximum achievable common throughput versus total transmit power for different algorithms with $M=3$.}
		\label{result2}
	\end{figure}
	
	Fig. \ref{result2} shows the maximum achievable common throughput $\bar{R}^*$ versus total transmit power $P_{\max}$ for different algorithms.
	We compare $\bar{R}^*$'s achieved by Algorithm 1, the low-complexity algorithm proposed in Section \ref{V-C}, and the ESM which finds the globally optimal antenna allocation via exhaustive enumeration.
	It is observed that Algorithm 1 achieves very close performance to the ESM, but with lower complexity. 
	Specifically, for each value of sector rotation $r$, the number of computations is in the order of $M=3$ for Algorithm 1 to determine the optimal antenna allocation, while the ESM requires evaluation over all possible antenna partitions, i.e., $44551$ combinations for $N=300$ and $M=3$.
	In addition, as $P_{\max}$ increases, the performance gap between Algorithm 1 and the low-complexity algorithm becomes larger.
	This is because, with more available transmit power, the performance becomes increasingly sensitive to the array gain and thus the antenna allocation. 
	Since the low-complexity algorithm does not jointly consider antenna allocation when optimizing sector rotation, its solution is suboptimal in high-power scenarios.
	Nevertheless, it still offers satisfactory performance with low complexity.
	We adopt Algorithm 1 for subsequent simulations due to its favorable tradeoff between performance and computational complexity.
	
	\begin{figure*}[t]	
		\centering
		\subfigure[Visual representation of the joint optimization results with $r^* = 6$, and the optimal antenna allocations  ${\bf n}^* ={\left[118,5,175\right]}^T$. ]{\includegraphics[width=0.48\textwidth]{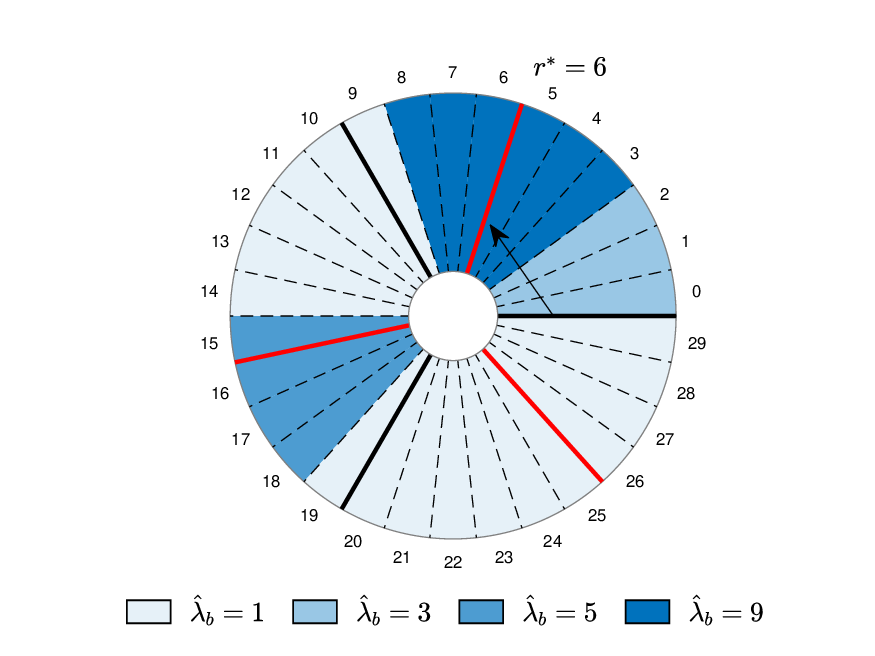}}
%		\hspace{0.05 in}
%		\subfigure[Achievable common throughput versus total transmit power for different algorithms.]{\includegraphics[width=0.32\textwidth]{one_realization_case_compare_algorithm.eps}}
		\hspace{0.05 in}
		\subfigure[Maximum achievable common throughput versus total transmit power for different schemes.]{\includegraphics[width=0.48\textwidth]{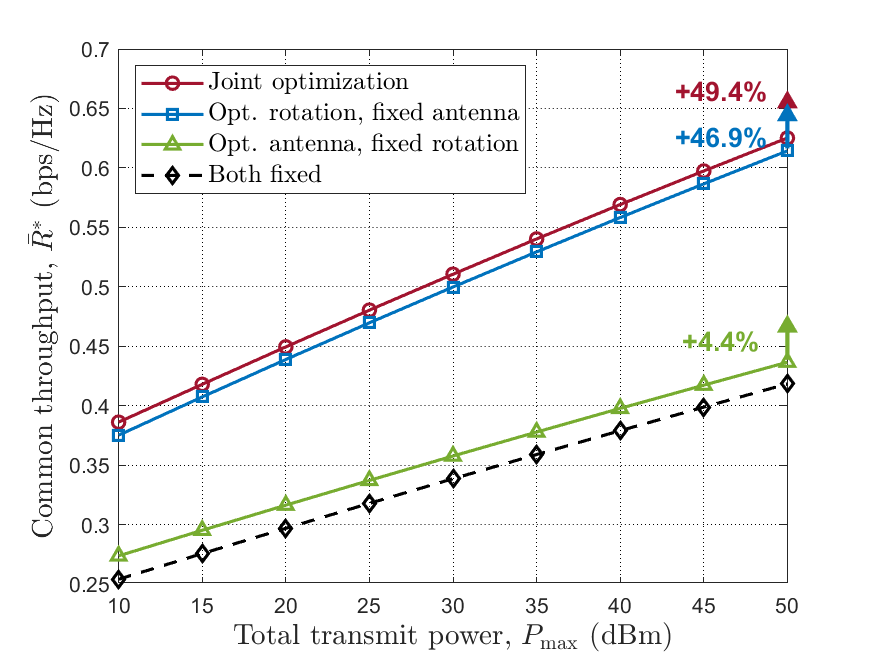}}
		\vspace{-4 pt}
		\caption{Performance comparison under a typical non-uniform user distribution scenario with $M=3$.% with fixed hotspot locations ${\left[\omega_{\mathrm{hot},1},\omega_{\mathrm{hot},2},\omega_{\mathrm{hot},3}\right]}={\left[0,3,15\right]} $.
		}\label{result3}
	\end{figure*}
	\subsection{Performance Evaluation with Non-uniform User Distribution}
	Fig. \ref{result3} shows a visual representation of the optimal sector rotation obtained by  Algorithm 1, and compares the maximum common throughput achieved by  the proposed joint optimization scheme with three benchmark schemes.
	In Fig. \ref{result3}(a), the black solid lines indicate the sector boundaries under the conventional fixed-sector BS  scenario. % with the fixed and equal antenna allocation and sector rotation $r=0$.
	The red solid lines correspond to the sector boundaries determined by the proposed joint optimization scheme.
	It is observed that, under the conventional setting, the user distribution across sectors is highly imbalanced with $\boldsymbol{\lambda}(0)=[64,26,10]^T$ and $\operatorname{Var}(\boldsymbol{\lambda}(0)) \approx 769 $. 
	In contrast, the optimal sector rotation effectively shifts the sector boundaries such that the average number of users across sectors becomes much more balanced with $\boldsymbol{\lambda}(6)=[38,22,40]^T$ and $\operatorname{Var}(\boldsymbol{\lambda}(6)) \approx 97 $. 
	This significant reduction in user distribution imbalance enables more efficient antenna allocation.
	As shown in Fig. \ref{result3}(b), the scheme with  antenna allocation optimization only provides marginal performance improvement over the fixed-sector BS scheme.
	This is because the exponentially growing power requirement for the sector with a large number of users limits system performance. 
	In this case, allocating additional antennas cannot sufficiently overcome this limitation, as the array gain remains constrained by the power budget.
	In contrast, the joint optimization scheme achieves the highest common throughput, demonstrating that adaptively adjusting sector rotation to balance user distribution is substantially more effective than solely optimizing antenna allocation when dealing with  the non-uniform user distribution in the case of OMA.
	
	\subsection{Average Performance Comparison with Benchmark Schemes}
	\begin{figure}[t]
		\centering
		\includegraphics[width=0.45\textwidth]{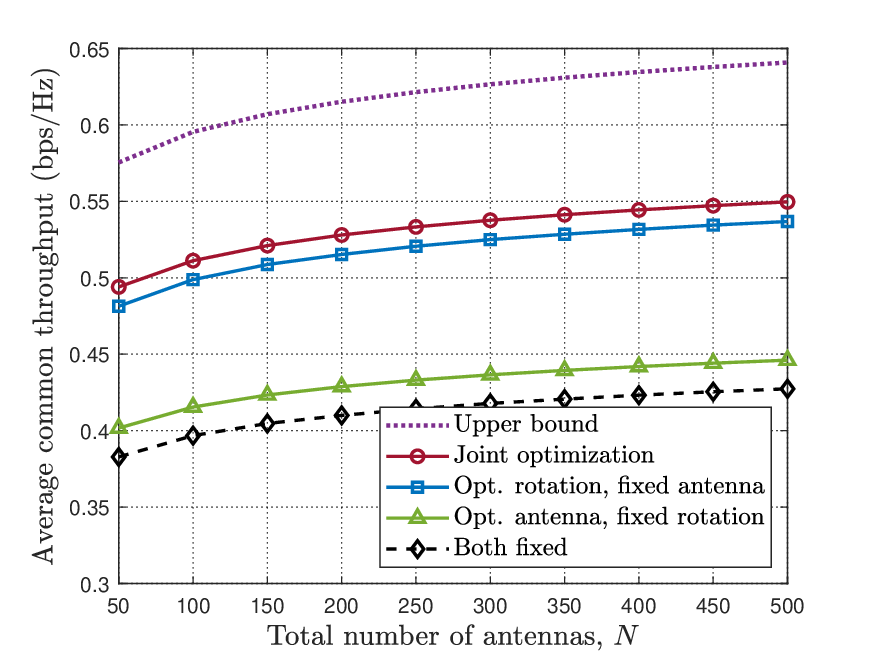}
		\vspace{-2 mm}
		\caption{Average achievable common throughput versus total number of antennas with $M=3$.}
		\label{result4}
	\end{figure}
	
	\begin{figure}[t]
		\centering
		\includegraphics[width=0.45\textwidth]{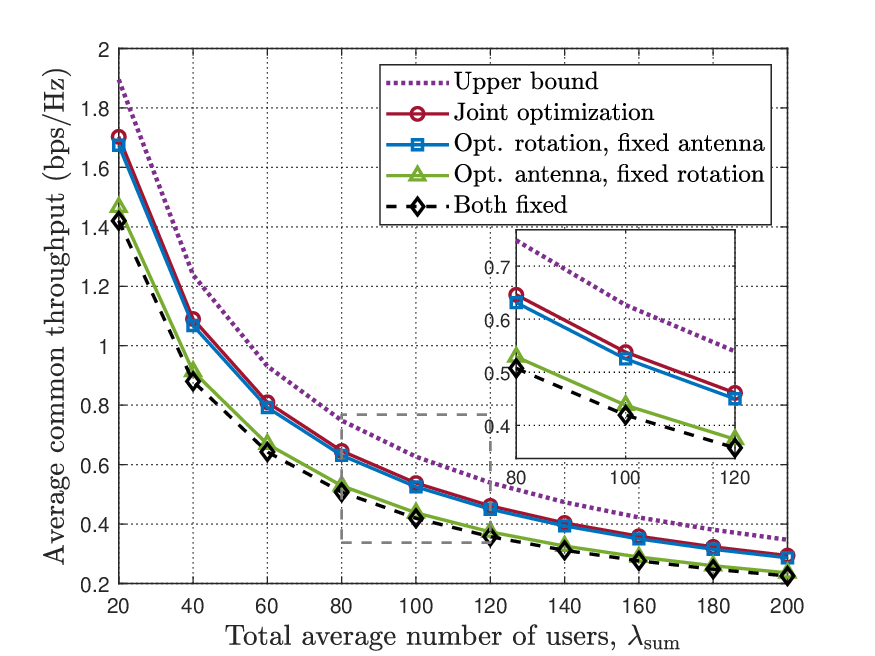}
		\vspace{-2 mm}
		\caption{Average achievable common throughput versus total average number of users with $M=3$.}
		\label{result5}
	\end{figure}
	
	In this subsection, to evaluate the effectiveness of the proposed schemes across random hotspot location configurations, we investigate the average achievable common throughput over $N_{\rm spl}=10^3$ independent realizations. 
	Specifically, for each realization, the starting bin indices of all hotspot areas $\omega_{\mathrm{hot},i},i=1,\cdots,N_\mathrm{hot},$  are independent and uniformly selected from the set of bins $\mathcal{B}$, i.e., $\omega_{\mathrm{hot},i} \sim \mathrm{Unif}\{0,1,\cdots,B-1\}$.
	The average achievable common throughput over all realizations is approximated by $\mathbb{E}[\bar{R}^*]\approx \frac{1}{N_{\rm spl}}\sum_{i=1}^{N_{\rm spl}}\bar{R}^*_i$, where $\bar{R}^*_i$ denotes the maximum achievable common throughput obtained in the $i$-th realization.
	
	Fig. \ref{result4} shows the average achievable common throughput versus the total number of antennas at the BS under different optimization schemes, along with the upper bound obtained by jointly optimizing arbitrarily adjustable user distribution and continuous antenna allocation, as given in \eqref{Rupperbound}.
    As observed, the proposed joint optimization scheme outperforms all benchmark schemes.
	However, there remains a performance gap between the joint optimization and the upper bound, due to  practical limitations in sector rotation for balancing the user distribution across sectors, as well as the loss resulting from discrete antenna allocation.
	In addition, the improvement in average achievable common throughput gradually saturates as the total number of antennas increases, due to the inherent logarithmic relationship between the throughput and the array gain.
	
	Fig. \ref{result5} shows the average achievable common throughput versus the total average number of users in the whole coverage area of the BS.
	It is observed that the average achievable common throughput decreases as the total average number of users increases for all schemes.
	This is because, as more users are served simultaneously, each user is allocated fewer  orthogonal resources in the case of OMA.
	To guarantee a target common throughput for all users, the required transmit power  increases  with the average number of users.
	Consequently, the system becomes increasingly constrained by the transmit power limit, resulting in a lower achievable common throughput.

	\begin{figure}[t]
		\centering
		\includegraphics[width=0.45\textwidth]{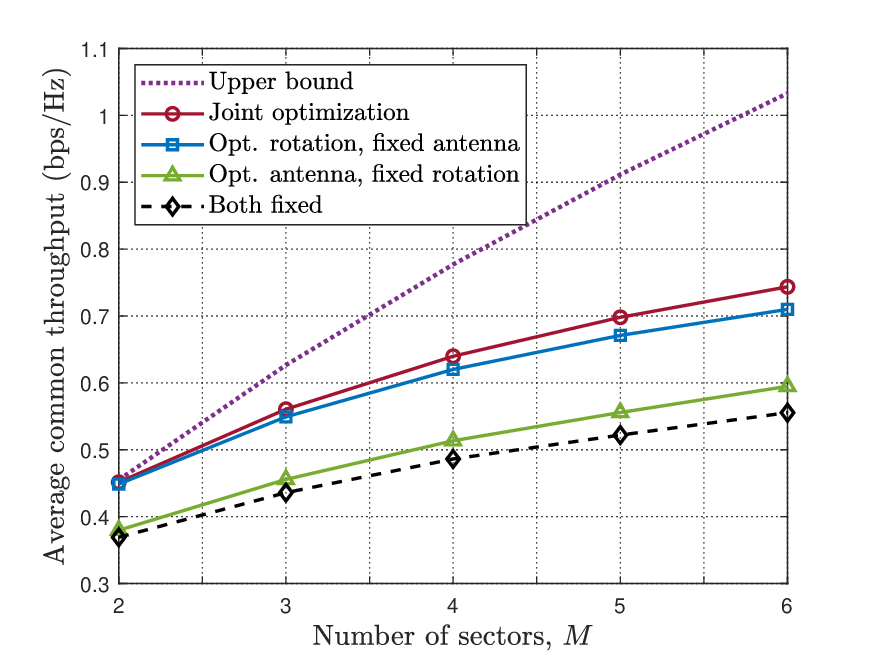}
		\caption{Average achievable common throughput versus total number of sectors with $B=60$.}
		\vspace{-2 mm}
		\label{result6}
	\end{figure}
	
	\begin{figure}[t]
		\centering
		\includegraphics[width=0.45\textwidth]{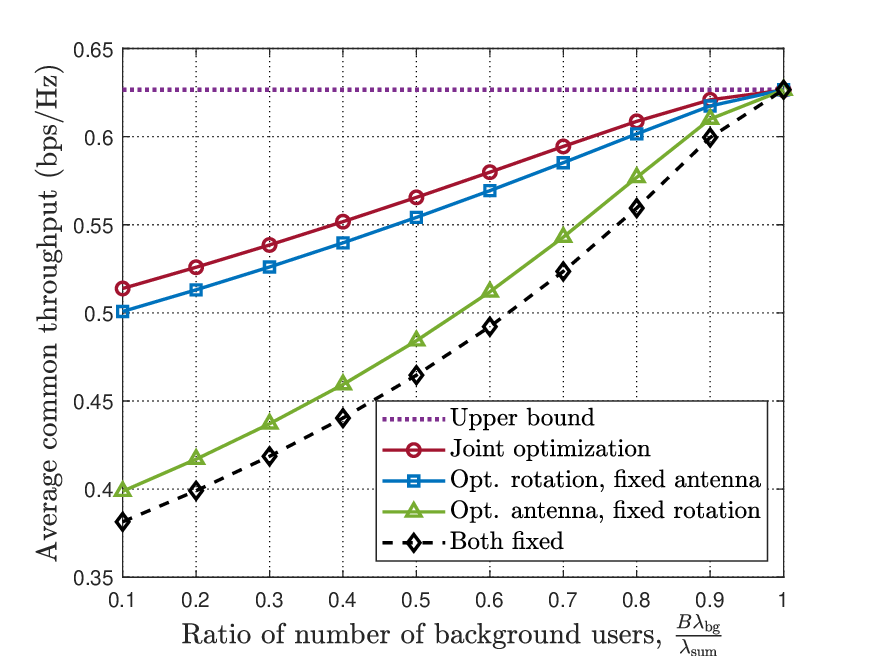}
		\vspace{-2 mm}
		\caption{Average achievable common throughput versus ratio of total number of background users with $M=3$.}
		\label{result7}
	\end{figure}
	
	Fig. \ref{result6} shows the average achievable common throughput versus the total number of sectors $M$, with the total number of bins fixed at $B=60$, to ensure a consistent rotation step for different values of $M$.
	It is observed that when $M=2$, the performance achieved by both the joint optimization scheme and the scheme with sector rotation optimization only is very close to the upper bound. 
	This is because, with $M=2$, it is much easier to achieve balanced user distribution across sectors via sector rotation.
	As $M$ increases, it becomes increasingly difficult for practical sector rotation to simultaneously achieve balanced user distribution across all sectors, leading to a growing performance gap between the practical schemes and the theoretical upper bound.
	In addition, the gap between the joint optimization scheme and the scheme with sector rotation optimization only also becomes more evident as $M$ increases. 
	This is because a larger number of sectors provides more DoFs for antenna allocation, thus enabling more effective compensation for residual user distribution imbalances that cannot be addressed by optimizing sector rotation only.
	
	In Fig. \ref{result7}, we investigate the effect of user distribution non-uniformity on the average achievable common throughput. 
	With the total average number of users fixed, it is observed that as the user distribution becomes more uniform, i.e., the ratio of the average number of background users to the total average number of users, $\frac{B\lambda_{\mathrm{bg}}}{\lambda_{\rm sum}}$, increases, the average achievable common throughput improves for all schemes. 
	When users are uniformly distributed across all sectors, i.e., $\frac{B\lambda_{\mathrm{bg}}}{\lambda_{\rm sum}}=1$,  all schemes achieve the same performance which nearly approaches that of the upper bound, with only a negligible gap remaining due to the loss resulting from discrete antenna allocation.
	These results confirm that, the optimal user distribution across sectors for the common  throughput maximization should lead to equal average numbers of users in all sectors.
	In addition, as the user distribution becomes more non-uniform, the performance gain provided by optimizing sector rotation over the conventional fixed-sector BS  scheme becomes more significant.

	\begin{figure}[t]
	\centering
	\includegraphics[width=0.45\textwidth]{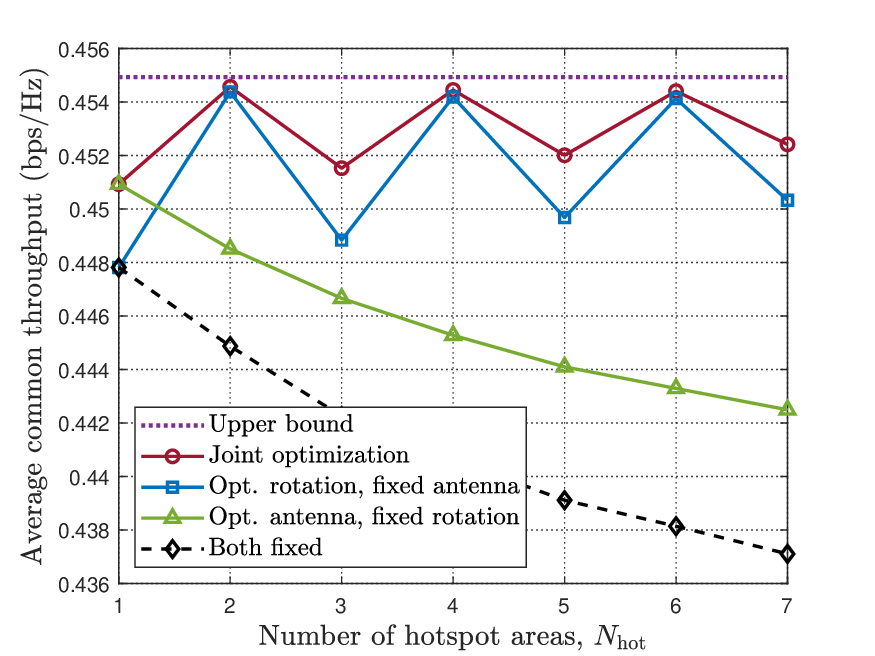}
	\vspace{-2 mm}
	\caption{Average achievable common throughput versus number of hotspot areas with $M=2$, $L=20$, ${{\hat\lambda}_{\mathrm{bg}}}:{{\hat\lambda}_{\mathrm{hot},i}}={1}:{2}$, and $L_{\mathrm{hot},i} = 1, \forall i$.}
	\label{result8}
	\end{figure}
	
	Fig. \ref{result8} illustrates the average achievable common throughput versus the number of hotspot areas $N_{\rm hot}$.
	For both the joint optimization scheme and the scheme with sector rotation optimization only, the average achievable common throughput fluctuates over $N_{\rm hot}$ and exhibits local peaks when $N_{\rm hot}$ is an integer multiple of the number of sectors $M$.
	This is because, in these cases, sector rotation can more evenly allocate hotspot areas across all sectors, thus leading to a more balanced and sometimes even nearly equal user distribution compared to cases where  $N_{\rm hot}$ hotspot areas cannot be evenly divided among all sectors.
	In contrast, for the schemes without sector rotation optimization, the average achievable common throughput decreases  as $N_{\rm hot}$ increases, as only optimizing  antenna allocation is insufficient to compensate for the increasing user distribution imbalance, resulting in  high performance degradation.
	
	\vspace{-2 mm}
	\section{Conclusion}\label{VII}
	In this paper, we introduced  a  flexible-sector BS design under the multiuser downlink communication setup.
	We first characterized the channel model between the flexible-sector BS and users served by different sectors, and derived the achievable user rate by considering the impacts of sector rotation and antenna allocation across sectors.
	Based on the knowledge of users' spatial distribution, we derived the closed-form expression for the total average transmit power required at each sector to achieve a given common throughput of all users.
	Then, we solved the common throughput maximization problem by jointly optimizing the sector rotation and the antenna allocation across sectors.
	Our study was further extended to the ideal setup with arbitrarily adjustable user distribution across sectors, where we optimized the numbers of users in all sectors jointly with, or under fixed, antenna allocation.
	We showed that equal user distribution over sectors is optimal for maximizing the common throughput and developed an alternative low-complexity suboptimal solution for the sector rotation by minimizing the variance of user numbers across sectors.
	Simulation results validated the accuracy of our analytical results and the effectiveness of the proposed algorithms.
	Compared to the conventional fixed-sector BS, the flexible-sector BS was shown to significantly improve the achievable common throughput by adjusting the user distribution across sectors via sector rotation  and allocating antennas over sectors via antenna movement.
	Furthermore, the results revealed that sector rotation adjustment is more effective  in enhancing throughput  compared to antenna movement in our considered setup.
	The flexible-sector BS can be broadly applicable to various wireless scenarios, e.g., multiuser sum-rate maximization systems \cite{li2025}, cell-free/distributed massive MIMO systems, and millimeter-wave/terahertz networks.

	\section*{Appendix A: Proof of Theorem 1}
	Based on \eqref{h_k}, \eqref{p_k}, and \eqref{barp_m}, the total average transmit power for sector $m$ is expressed as
	\begin{equation}\label{exp_p_m}
		\begin{aligned}
			\bar{p}_m(r,N_m,\bar{R})& = \frac{\sigma^2}{\tau M N_m} \mathbb{E}_{\mathcal{P}_m}\bigg[\frac{2^{{K_m(r)}\bar{R}}-1}{{K_m(r)\beta_0}}\\
			&\quad\ \times\sum_{k\in\mathcal{K}_{m}(r)} \left(d_k^2+H^2\right)^{\frac{\alpha_0}{2}}\bigg]\\
			&\overset{(a)}{=}\frac{\sigma^2}{\tau M N_m} \mathbb{E}_{K_m(r)}\bigg[\mathbb{E}_{\bf d}\bigg[\frac{2^{{K_m(r)}\bar{R}}-1}{{K_m(r)\beta_0}}\\
			&\quad\ \times\sum_{k\in\mathcal{K}_{m}(r)} \left(d_k^2+H^2\right)^{\frac{\alpha_0}{2}}\left|  K_m(r)  \right.\bigg]\bigg],
		\end{aligned}
	\end{equation}
	where ${\bf d}$ denotes the collective vector of horizontal user distances from the BS, $d_k,k\in\mathcal{K}_m(r)$, and step $(a)$ follows from the law of total expectation.
	
	For the distance distribution, as the users are independent and uniformly distributed within each bin, the CDF of the horizontal distance from user $k\in\mathcal{K}$ to the BS's reference position is calculated as the area of the ring with inner radius $D_{\min}$ and outer radius $D_{\max}$, divided by the area of the ring with inner radius $D_{\min}$ and outer radius of the user's  horizontal distance, given by \vspace{-1 mm}
	\begin{equation}\label{cdf_d}\vspace{-1 mm}
		F(d) = \left\{ {\begin{array}{*{20}{l}}
				{ \frac{d^2-D_{\min}^2}{D_{\max}^2-D_{\min}^2} ,}&{D_{\min}\leq d \leq D_{\max},} \\ 
				0 ,&{\text{otherwise.}} 
		\end{array}} \right.
	\end{equation}
	By taking the derivative of CDF $F(d)$ w.r.t. $d$, the probability density function (PDF) of the horizontal distance from user $k\in\mathcal{K}$ to the BS's reference position is obtained as \vspace{-1 mm}
	\begin{equation}\label{pdf_d}\vspace{-1 mm}
		f(d) = \left\{ {\begin{array}{*{20}{l}}
				{ \frac{2d}{D_{\max}^2-D_{\min}^2} ,}&{D_{\min}\leq d \leq D_{\max},} \\ 
				0 ,&{\text{otherwise.}} 
		\end{array}} \right.\vspace{-2pt}
	\end{equation}
	Thus, given $K_m(r) = n$, the inner expectation over $\bf d$ in \eqref{exp_p_m} is given by
	\begin{equation}\label{exp_d}
		\begin{aligned}
			\mathbb{E}_{\bf d}&\bigg[\frac{2^{{n}\bar{R}}-1}{{n}\beta_0}\sum_{k=1}^n \left(d_k^2+H^2\right)^{\frac{\alpha_0}{2}}\bigg]\\
			&\overset{(b)}{=} \frac{2^{{n}\bar{R}}-1}{\beta_0}\mathbb{E}_{d_k}\bigg[ \left(d_k^2+H^2\right)^{\frac{\alpha_0}{2}}\bigg]\\
			&=\frac{2({2^{{n}\bar{R}}-1})}{(D_{\max}^2-D_{\min}^2)\beta_0} \int_{D_{\min}}^{D_{\max}} \left(d_k^2+H^2\right)^{\frac{\alpha_0}{2}}d_k~{\rm d}d_k\\
			& = ({2^{{n}\bar{R}}-1})\nu_0,
		\end{aligned}\vspace{-2pt}
	\end{equation}
	where step $(b)$ follows from the fact that the users are independent and uniformly distributed in the sector.
	Substituting \eqref{exp_d} into \eqref{exp_p_m}, we have\vspace{-1 mm}
	\begin{equation}\vspace{-1 mm}
		\begin{aligned}
			\bar{p}_m(r,N_m,\bar{R})& = \frac{\nu_0\sigma^2}{\tau M N_m} \mathbb{E}_{K_m(r)}\left[2^{{K_m}\bar{R}}-1 \right]\\
			& \overset{(c)}{=} \frac{1}{\gamma_m(N_m)}e^{\bar{\gamma}\lambda_m(r)},
		\end{aligned}
	\end{equation}
	where step $(c)$ follows from the moment generating function (MGF) of the Poisson distribution of $K_m(r)$ with $t=\bar{R}\ln 2$, which is given by\vspace{-1 mm}
	\begin{equation}\label{MGF}\vspace{-1 mm}
		\begin{aligned}
			{\rm MGF}_{K_m}(t) &= \mathbb{E}\left[e^{tK_m(r)}\right] \\
			&= e^{\lambda_m(r)\left(e^t-1\right)},\quad m\in\mathcal{M}.
		\end{aligned}
	\end{equation}
	This thus completes the proof of Theorem \ref{thm1}.\vspace{-1 mm}
	
	\bibliographystyle{IEEEtran}
	\bibliography{123}
\end{document}